\documentclass[prx,superscriptaddress,twocolumn,nofootinbib]{revtex4-2}

\usepackage{graphicx,epsfig}
\usepackage{color}
\usepackage[usenames,dvipsnames]{xcolor}
\usepackage{amsmath,bbm,amssymb, amsthm}
\usepackage{dsfont} 
\usepackage{stmaryrd}
\usepackage{comment}
\usepackage{hyperref}
\usepackage{numprint}

\makeatletter
\newcommand{\fmarki}{\ensuremath{\dagger}}
\newcommand{\fmarkii}{\ensuremath{\ddagger}}
                
\def\@fnsymbol#1{{\ifcase#1\or \fmarki\or \fmarkii \else\@ctrerr\fi}}
\makeatother

\def\bb{\begin{eqnarray}}
\def\ee{\end{eqnarray}}
\newcommand{\ket}[1]{| #1 \rangle}
\newcommand{\bra}[1]{\langle #1 |}

\newcommand{\moy}[1]{\left\langle #1 \right\rangle} 
\newcommand{\idop}{\mathds{1}}

\renewcommand{\>}{\rangle}

\newcommand{\nn}{\nonumber}

\begin{document}

\title{Extending the laws of thermodynamics for autonomous, arbitrary quantum systems}

\author{Cyril Elouard$^{*}\,$}
\email{cyril.elouard@gmail.com}

\affiliation{Inria, ENS Lyon, LIP, F-69342, Lyon Cedex 07, France}

\author{ Camille Lombard Latune$^{*}\,$ }
\email{cami-address@hotmail.com}
\thanks{\\$^{*}$ These two authors contributed equally.}
\affiliation{ENSL, CNRS, Laboratoire de physique, F-69342 Lyon, France}

\begin{abstract}
Originally formulated for macroscopic machines, the laws of thermodynamics were recently shown to hold for quantum systems coupled to ideal sources of work (external classical fields) and heat (systems at equilibrium). Ongoing efforts have been focusing on extending the validity of thermodynamic laws to more realistic, non-ideal energy sources. Here, we go beyond these extensions and show that energy exchanges between arbitrary quantum systems are structured by the laws of thermodynamics. We first generalize the second law and identify the associated work and heat exchanges. After recovering known results from ideal work and heat sources, we analyze some consequences of hybrid work and heat sources. We illustrate our general laws with microscopic machines realizing thermodynamic tasks in which the roles of heat and work sources are simultaneously played by elementary quantum systems. Our results open perspectives to understand and optimize the energetic performances of realistic quantum devices, at any scale.
\end{abstract}

\maketitle

\begin{figure*}
\begin{center}
    \includegraphics[width=0.7\textwidth]{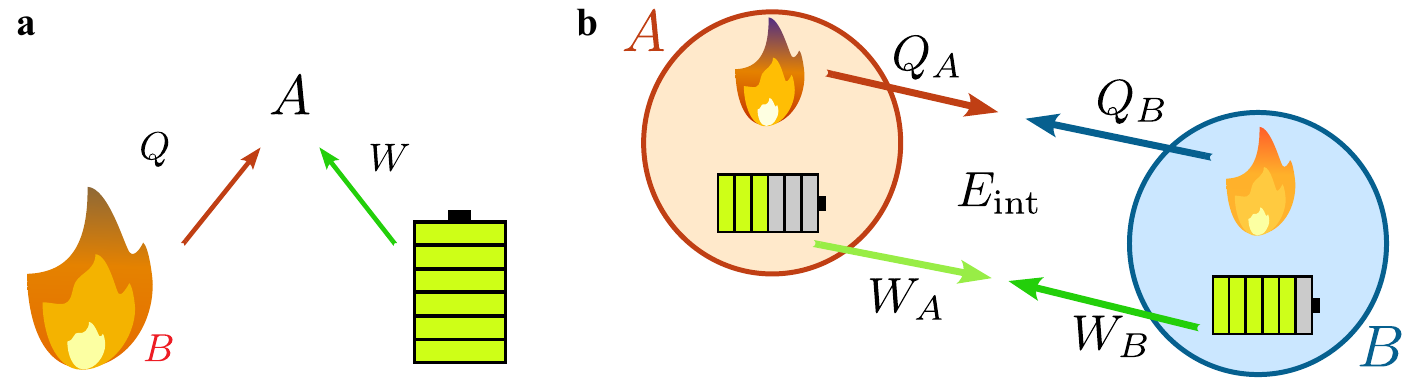}
\end{center}
\caption{\textbf{a}: Recent formulations of the laws of thermodynamics apply to a system $A$ (which can be microscopic and/or quantum), coupled to ideal (macroscopic) sources of heat $Q$ and of work $W$. \text{b}: We demonstrate the laws of thermodynamics for two or more quantum systems, that can be of any scale and initialized in any arbitrary uncorrelated state.
Each system can in general behave as both a source of work and heat to the other system. Note that the heat (respectively work) provided by system $A$ does not necessarily equals the heat (work) received by system $B$ as some energy $E_\text{int}$ can be stored in the coupling between the systems and as one kind of energy can be consumed to the profit of the other one.}\label{f:1}
\end{figure*}

\section{Introduction}
The laws of thermodynamics have been formalized centuries ago to predict the performances of macroscopic machines, composed of large systems exchanging work and heat. While the first law of thermodynamics defines the splitting of the energy exchanges between heat and work, the second law states their different nature by expressing the fundamental constraints they obey due to the arrow of time.
In the last two decades, the application domain of these concepts has been dramatically expanded, as similar laws were found to rule the average energy flows received by a single nanoscopic classical \cite{Jarzynski97,Crooks99,Jarzynski11,Seifert12}, or quantum system \cite{Spohn78,Alicki79,Kurchan00,Esposito10,Hasegawa10,QTDBook,Landi21,Strasberg21}, far from equilibrium and from the so-called thermodynamic limit. 
Importantly, the second law was shown to emerge universally from the Schr\"odinger equation ruling the dynamics of a quantum system of interest coupled on the one hand to a heat bath -- a quantum system initially at thermal equilibrium and identified with a pure source of heat -- and on the other hand to a pure source of work \cite{Esposito10}  -- modeled as a time-dependence in the system's Hamiltonian. 
Recent advances in the analysis of quantum effects in heat engines has however motivated to look further into full quantum description of sources of heat and work \cite{Gelbwaser14,Skrzypczyk14,Elouard15,Brandao15,Alhambra16,Muller18,Seah18,Bera19,Strasberg21,Jacob21,Jacob22}. 
Indeed, time-dependent Hamiltonians can be considered as effective semi-classical models. One could expect that work exchanges should emerge from a fully autonomous scenario, where the system, the sources of work, and the sources of heat are described by a global time-independent Hamiltonian. In such scenarios, the division between heat and work becomes fuzzy, rising doubt whether a consistent thermodynamic framework can be built at this level. Some noticeable progresses in that direction include the introduction of the notion of ergotropy \cite{Pusz78,Allahverdyan04} as work contribution in fully autonomous systems \cite{Gelbwaser13,Gelbwaser14,Ghosh17,Seah18,Uzdin18,Uzdin19,Mitchison2019} and extensions of the traditional frameworks to non-thermal baths \cite{Manzano16,Santos17,Niedenzu18,Sparaciari17,Bera19,Macchiavello2020}.
In parallel to these developments, it has been noticed that energy exchanges with a large quantum system initially at thermal equilibrium, which would traditionally be regarded as heat source, can also contain coherent, deterministic contributions, exhibiting the properties of work \cite{Uzdin19,Monsel20,Bernhardt22}.
 As a matter of fact, in several implementations of quantum heat engines, the very same physical devices can often be used to provide either work or heat (e.g. a microwave transmission line can both induce thermalization or convey a driving field performing work on a superconducting qubit \cite{Nathanael17}), reinforcing the need to describe hybrid quantum energy sources providing both work-like and heat-like energy.
All the above efforts contributed to build fundamental definitions of work and heat in the quantum regime which do not presuppose the role of work or heat sources for each device, but relate directly to the properties of the exchanged energy \cite{Weimer08,Hossein15,Maffei21}.

Our motivation here is to go beyond the dichotomy of pure heat and work sources/storages used in many frameworks \cite{Esposito10,Brandao13,Skrzypczyk14,Alhambra16,Elouard20,Strasberg21}, and rather describe each quantum system on equal footing as an hybrid source of heat and work. Building on the above studies, we introduce an expression of the second law of thermodynamics valid for arbitrary set of interacting quantum systems, which leads us to identify consistent definitions for heat and work. 
We show that the emerging notion of work corresponds to a generalization of the concept of ergotropy, the latter being in general insufficient to fully quantify energetic resources that can be consumed to decrease entropy, as we show on specific examples. 
On the other hand, the notion of heat which naturally emerges is the variation of complete passivity \cite{Alicki13,Sparaciari17,Bera19}, which, interestingly, sets an intrinsic notion of temperature as developed in \cite{Sparaciari17,Bera19,Macchiavello2020}.

As an illustrative application of the suggested framework, we show that sets of quantum systems as small as two interacting qubits can behave as autonomous thermal machines, where work is consumed to cool down one of the qubit (refrigerator), or conversely, where work is produced out of initial temperature gradient between the two qubits (engine). We show that the efficiencies of both operations are limited by the same upper bounds as classical macroscopic thermal machines, namely the Carnot bounds. 
Finally, we explain how to use this framework to analyze more complex autonomous machines, including $N$ interacting systems, with autonomous switching of couplings Hamiltonians.

\section{Background: Second law for a quantum system coupled to ideal heat and work sources}

Before stating our results, we recall one of the most general settings demonstrating the emergence of the second law from unitary global evolution.
We consider a quantum system $A$ that is externally driven and interacts with a heat bath, modeled by a global Hamiltonian of the form $H(t) = H_A(t) + V(t) + H_B$, where $V(t)$ contains the system-bath interaction terms. The action of the drive is captured semi-classically by the time-variation of $H(t)$.
We assume that at time $t=0$, the system $A$ and the bath $B$ are in a factorized state $\rho_S(0)\otimes w_B[\beta_B]$, where $w_B[\beta_B] = e^{-\beta_B H_B}/Z_B$ denotes the thermal equilibrium state at inverse temperature $\beta$, $Z_B$ being the partition function. At times $t>0$, the joint dynamics of the two systems generates a correlated state of both systems $\rho_{AB}(t)$, exhibiting in general a non-zero mutual information, which can be associated with a production of entropy. More precisely, the following equality has been derived \cite{Esposito10}:
\bb
\sigma_0(t) &=& \Delta S_A(t) + \beta_B \Delta E_B(t)\nonumber\\ &=& I_{SB}(t)+ D(\rho_B(t)\vert\rho_B^\text{eq}). \label{2ndLaw_1}
\ee
Here, $\Delta X(t) = X(t) - X(0)$ for any quantity $X$. Moreover, $S_A = S[\rho_A(t)]$ denotes the Von Neumann entropy of system $A$, with $S[\rho] = -\text{Tr}\{\rho\log\rho\}$, and $D(\rho_1\vert\vert \rho_2) = \text{Tr}\{\rho_1(\log \rho_1 -\log \rho_2)\}$ denotes the relative entropy of states $\rho_1$ and $\rho_2$. Throughout the article, we use natural units such that $k_B=\hbar = 1$. We have also introduced the partial states of the system $\rho_A = \text{Tr}_B\{\rho_{AB}(t)\}$ and the bath $\rho_B = \text{Tr}_A\{\rho_{AB}(t)\}$. In addition $I_{AB}(t) = D(\rho_{AB}(t)\vert\vert \rho_A(t)\otimes\rho_B(t))$ is the mutual information of $A$ and $B$ that built up  during their joint evolution ($I_{AB}(0)=0$). As the mutual information and the relative entropy are positive quantities, so is the right-hand side of Eq.~\eqref{2ndLaw_1}. As a consequence, $\sigma_0$ is often interpreted as the second law of thermodynamics for quantum systems \cite{Deffner11,Jarzynski11,Goold16,Campisi16,Bedingham16}, and the energy change of the reservoir is usually categorized as heat (interpreting $-\beta_B \Delta E_B$ as the entropy exchanged with the reservoir). Introducing the work performed by the external drive as the change of energy of the total system $A+B$, namely $W_\text{dr}(t) = \int_0^t dt' \text{Tr}\{\dot H(t')\rho_{AB}(t')\}$, and defining the nonequilibrium free energy of system $A$, namely $F_A(t) = E_A(t)-S_A(t)/\beta_B$, one can also rewrite Eq.~\eqref{2ndLaw_1} as:
\bb
W_\text{dr}(t)\geq \Delta F_A + \Delta E_\text{int},\label{2ndLaw_F1}
\ee
with $E_\text{int}(t) = \moy{V(t)}$ the energy stored in the coupling. We recover in this way a usual statement of the second law for an isothermal transformation leading to a change of free energy $\Delta F_A$, which takes into account the role of the interaction energy when the latter cannot be neglected (as it is often the case for nanoscale systems).
Finally, we stress that in this paradigm, the irreversibility of an evolution, as quantified by the entropy production, is ultimately related to lack or loss of information occurring when tracing out the bath $B$. \\

\section{Second law for two arbitrary quantum systems}\label{s:SecondLawAB}

We now present our framework allowing to extend the second law to two arbitrary quantum systems $A$ and $B$ initially in arbitrary local states (i.e. an arbitrary uncorrelated state $\rho_A(0)\otimes\rho_B(0)$ of the joint system), and which can then be sources of both work and heat. The difference with the above setup is therefore that we now allow the state of $B$ to be initially not in a thermal state. For the sake of completeness and pedagogy, we still allow for the Hamiltonian of system $A$ to be explicitly time-dependant in Sections \ref{s:SecondLawAB}-\ref{s:PropWork}, but our final goal is to consider time-independent Hamiltonians, and show the autonomous emergence of the notion of work. Moreover, we wish to separate the energy exchanges between the systems into in heat-like contributions that will be related to entropy and a work-like part that can be considered as a resource for instance to decrease entropy, consistently with the expression of the second law.

This splitting can be introduced by comparing the state $\rho_B(t)$ of system $B$ at any time $t$ to the thermal state $w_B[\beta_B(t)] = e^{-\beta_B(t)H_B}/Z_B(t)$ which is chosen to have the same entropy as $\rho_B(t)$. Here $Z_B(t) = \text{Tr}\{e^{-\beta_B(t)H_B}\}$ is the associated partition function. Namely, the effective inverse temperature $\beta_B(t)$ is defined via the equation:
\bb
S[w_B[\beta_B(t)]] =  S[\rho_B(t)], 
\ee
where the left-hand term is a function of $\beta_B(t)$ only $S[w_B[\beta_B(t)]]=\beta_B(t)\text{Tr}\{H_B w_B[\beta_B(t)]\} +\log Z_B(t)$. As $S[w_B[\beta_B(t)]$ is a monotonously decreasing function of $\beta_B(t)$ spanning the interval $[0,\log d_B]$ of all the possible values taken by $ S[\rho_B(t)]$ ($d_B \in[2;+\infty)$ is the dimension of the Hilbert space of $B$), this equation always admits a unique solution, defining uniquely the thermal state $w_B[\beta_B(t)]$. 
Note that this appealing intrinsic notion of temperature was extensively analyzed in \cite{Sparaciari17,Bera19,Macchiavello2020}. We now introduce the quantity \cite{Bera19,Macchiavello2020}:
\bb
E_B^\text{th}(t) = \text{Tr}\{H_Bw_B[\beta_B(t)]\},
\ee 
that we refer to in the following as the \emph{thermal energy} of $B$, so as to state our central result:
\bb
\sigma_A &\equiv&\Delta S_A(t) - \beta_B(0)Q_B(t)\nonumber\\
&&= I_{AB}(t)+ D(w_B[\beta_B(t)]||w_B[\beta_B(0)]) \geq 0, \label{2ndLaw_AB}
\ee
with 
\bb\label{d:heat}
Q_B(t) =  -\Delta E_B^\text{th}(t).
\ee
\emph{Proof} -- Using that the unitary evolution of $A$ and $B$ preserves the total Von Neumann entropy, we can write:
\bb
\Delta S_A  = I_{AB}(t)-\Delta S_B.
\ee
Adding on both sides $\beta_B(0)\Delta E_B^\text{th}$ and using that
\bb
&&\beta_B(0)\Delta E_B^\text{th} - \Delta S_B\nonumber\\
&=& -\text{Tr}\{(w_B[\beta_B(t)]-w_B[\beta_B(0)])\log w_B[\beta_B(0)]\}\nonumber\\
&&- S[w_B[\beta_B(t)]] + S[w_B[\beta_B(0)]]\nonumber\\
&=& D(w_B[\beta_B(t)]\|w_B[\beta_B(0)]]),\ee
where to go to the second line we have used that by definition $S[w_B[\beta_B(t)]] = S_B(t)$, we finally obtain Eq.~\eqref{2ndLaw_AB}~$\blacksquare$\\

We interpret Eq.~\eqref{2ndLaw_AB} as the second law of thermodynamics for a transformation of system $A$ caused by the interaction with system $B$. By comparing with Eq.~\eqref{2ndLaw_1}, we identify $Q_B(t)$ as the heat \emph{provided} by system $B$ during the transformation of $A$ (with positive sign when the heat exits system $B$).  In addition, we observe the emergence of an \emph{effective inverse temperature} $\beta_B(0)$ associated with the initial entropy of system $B$, which sets constraints on this heat flow via Eq~\eqref{2ndLaw_AB}. We stress that system $B$ itself is not in thermal equilibrium (and a consequence will be that it can also provide work), but the definition of heat compatible with Eq.~\eqref{2ndLaw_AB} is the energy difference between the two ``thermal backgrounds" $w_B[\beta_B(t)]$ and $w_B[\beta_B(0)]$ associated with $\rho_B(t)$ and $\rho_B(0)$. Thus, $Q_B(t)$ corresponds to an energy exchange intrinsically linked to entropy change, verifying notably 
\bb
\dot{S}_B(t) = -\beta_B(t)\dot Q_B(t).\label{eq:dSbetaQ}
\ee

We note that the second law for system $A$, Eq.~\eqref{2ndLaw_AB}, involves the initial effective temperature of $B$. As it can be inferred from Eq.~\eqref{eq:dSbetaQ}, this effective temperature will change in general during the process. We show in Section \ref{s:Generalizations} that a tighter bound on $\Delta S_A$ can be found when keeping track of this time-dependence.
\\

\section{Properties of work}\label{s:PropWork}

Simultaneously to behaving as a heat source, system $B$ also behaves as a source of work for $A$. The quantity, 
\bb
W_B = -\Delta E_B - Q_B 
\label{FirstLawB},
\ee
plays the role of the work performed by system $B$. 
This can be illustrated by introducing the nonequilibrium free energy of system $A$ at temperature $\beta_B(0)$, namely $F_A[\beta_B(0)] = E_A(t)-S_A/\beta_B(0)$, to obtain from Eq.~\eqref{2ndLaw_AB}:
\bb
W_\text{dr}(t)+W_B \geq \Delta F_A[\beta_B(0)] +\Delta E_\text{int}(t).\label{2ndLaw_FA}
\ee
We see that $W_B$ and the external driving work $W_\text{dr}(t)$ appear on an equal footing in Eq.~\eqref{2ndLaw_FA} as resources that can be consumed to vary the system's free energy. Moreover, in the absence of external driving (i.e. a fully autonomous machine described by a time-independent Hamiltonian), one can identify the work $W_B$ spent by system $B$ without resorting to a semi-classical description. 
In what follows, we will always assume a time-independent total Hamiltonian such that $W_\text{dr}=0$. We also stress that during a transformation where $\Delta S_B = 0$, $W_B$ corresponds to the whole energy variation of system $B$, as expected. This justifies the intuition that work is the iso-entropic part of the energy exchange.  \\

Finally, we emphasize that 
the quantities $Q_B$ and $W_B$ only depend on the initial and final states of system $B$, not on its whole trajectory. Nevertheless, these quantities depend on the trajectory of $A$. Indeed, in an autonomous scenario where the total Hamiltonian is time-independent, the trajectory of $A$ can be varied only by selecting different parameters for $B$ (varying $H_B$, $V$ or $\rho_B(0)$), which in general will affect the final state of $B$ and therefore the value of $Q_B$ and $W_B$. 
We therefore retrieve the expected properties from standard non-autonomous machines, namely that the heat and work provided by the sources depend on the trajectory of $A$. An infamous consequence is that less work can be extracted in a stroke if the transformation is performed faster. This point is illustrated within our formalism on the case of a two-qubit engine (see Section \ref{s:qubitengine}).

\section{Work-like resources beyond erogtropy}

It is interesting to note that the definition of work $W_B$ which emerges from our formalism is connected to the concept of ergotropy, often considered as the adequate notion of work 
when energy sources are described quantum-mechanically \cite{Pusz78,Allahverdyan04,Gelbwaser13,Gelbwaser14,Seah18,Niedenzu18}. The ergotropy ${\cal E}_B$ of system $B$ with Hamiltonian $H_B$ is defined as the maximum amount of energy that can be extracted via a unitary transformation on $B$. There is an extended notion of ergotropy, defined as  ${\cal E}_B^\infty =  E_B -  E_B^\text{th}$, which was introduced as the maximal amount of energy which can be extracted from an infinite number of copies of the system via global unitary operations on all the copies \cite{Alicki13} (see \cite{Sparaciari17} for alternative extraction protocol). The notion of work emerging from the second law Eq.~\eqref{2ndLaw_AB} is precisely the variation of  ${\cal E}_B^\infty$.
This quantity ${\cal E}_B^\infty$ also bears two other important physical interpretations \cite{Bera19}: (i) it is the minimum amount of work one needs to pay to prepare the nonequilibrium state of $B$ from thermal equilibrium, with the help of a thermal bath (see Appendix A) and (ii) it is the maximum amount of energy that can be extracted during a transformation preserving the entropy.\\

The inequality ${\cal E}_B^\infty \geq {\cal E}_B$ \cite{Alicki13} implies that there exist other resources than the ergotropy which can be consumed to obtain effects similar to a work expense (that is, e.g. inducing a heat flow from a cold to a hot system, or reducing a system's entropy). This result can be made intuitive by noting that a state with zero ergotropy -- a so-called passive state -- is diagonal in the energy eigenbasis, with populations decreasing as a function of energy. When this state is not a thermal state, the ratio of these populations takes the form $p_{k+1}/p_k = e^{-\beta_k(E_{k+1}-E_k)}$ as a function of the energies $E_{k}$ with positive numbers $\beta_k$ which depend on $k$. These numbers can be interpreted as \emph{local} temperatures, whose biases can be exploited to perform work, provided one can access such state locally in energy, i.e. couple selectively to its level transitions exhibiting different values of $\beta_k$. This mechanism is illustrated in Section \ref{S:ExtractionPassive}.

Reciprocally, using the variation of ergotropy instead of the variation of ${\cal E}_B^\infty$ would lead to breaking down the positivity of the entropy production in some cases. In Appendix B, we show a concrete example based of two qutrits, demonstrating the need to identify the variation of ${\cal E}_B^\infty$ as work rather than the variation of ${\cal E}_B$. A necessary condition to obtain such a positivity violation is to choose an initial state verifying ${\cal E}_B^\infty \neq {\cal E}_B$, which is possible only in dimension greater than 2.

\section{ General consequences of Eq.~(5)}

\subsection{A tighter bound for a bath starting at equilibrium}

First, we consider the case where system $B$ is initially in a thermal state $w_B[\beta_B(0)]$ (which is when Eq.~\eqref{2ndLaw_1} applies). In this case, $\sigma_A$ still differs from $\sigma_0$ as our approach identifies a work contribution to the variation of the energy of $B$ when its final state is out-of-equilibrium. As $E_B(t)\geq  E_B^\text{th}(t) $, 
we have $\sigma_A\leq \sigma_0$, meaning that Eqs.~\eqref{2ndLaw_AB} and \eqref{2ndLaw_FA} correspond to tighter constraints than Eq.~\eqref{2ndLaw_1}-\eqref{2ndLaw_F1} on the entropy variation of $A$ and on the work cost required to perform a given transformation (associated with a given free energy variation).

\subsection{A fully symmetric description}

A second important consequence of our formulation is that system $A$ and $B$ can be treated on an equal footing. We define the entropy production $\sigma_B$ from the point of view of system $B$ by swapping the roles of $A$ and $B$ in Eq.~\eqref{2ndLaw_AB}. In general $\sigma_A\neq\sigma_B$, reflecting the fact that tracing over one or the other system does not lead to the same information loss. 
We can yet obtain an expression emphasizing the symmetric roles of the two systems by rewriting in Eq.~\eqref{2ndLaw_AB} the entropy variation of system $A$ as $\Delta S_A = -\beta_A(0)Q_A(t) - D(w_A[\beta_A(t)]\|w_A[\beta_A(0)])$, leading to (see Appendix C): 
\bb
-\beta_B(0)Q_B(t)-\beta_A(0)Q_A(t)\geq 0.\label{Clausius}
\ee
Eq.~\eqref{Clausius} holds even though each system also may also provide an amount of work $W_i(t) = -\Delta E_i(t) + \Delta E_i^\text{th}(t)$. It therefore expresses a universal constraint on the thermal part of the energy exchanges between systems $A$ and $B$ which is especially precious to express the limitations on nanoscale machine performances (see e.g. Section \ref{s:bipartitem}).
Eq.~\eqref{Clausius} bears similarities with the results presented in \cite{Niedenzu18}, obtained by focusing on the dynamics of the system only, and for the special case where $B$ is a bath initially in a state unitarily related to a thermal state. By including explicitly the system playing the role of heat and work source, we were able to reach a more general law applicable at any scale.

 \subsection{Accessing the effective temperature}\label{s:accessefft}

The price to pay to encompass any initial state of the systems and quantify any work-like resource is that more details about the reduced dynamics must be tracked to  express Eq.~\eqref{2ndLaw_AB} than Eq.~\eqref{2ndLaw_1}. Specifically, the initial and final effective temperatures of system $B$ need to be known, which requires the ability to either determine its von Neumann entropy (e.g. from a tomography protocol) or to simulate its evolution. While this may be challenging for large systems, it is possible with some theoretical methods \cite{Hartmann17} or experimental setups \cite{Eichler11}, e.g. allowing to track the Husimi function of a bath and therefore reconstruct its state \cite{Mizrahi84,Landon-Cardinal18M}.

We mention that a different notion of nonequilibrium temperature has been proposed, based on the thermal state which shares the same {\it energy} as $\rho_B(t)$; explicitly, it corresponds to the inverse temperature $\beta^*_B(t)$ verifying $\text{Tr}\{H_B w_B[\beta^*_B(t)]\}=\text{Tr}\{H_B \rho_B(t)\}$ \cite{Strasberg21}. 
 Contrary to the entropy-based temperature we employ here, $\beta^*_B(t)$
 can be negative for finite systems in non-passive states. 
 Note also that the entropy-based inverse temperature $\beta_B(t)$ is always smaller than $|\beta_B^*(t)|$.
 Instead of identifying resources able to produce work in the state of $B$,
 the energy-based temperature treats all its energy changes as heat.  
 On an operational level, the energy-based temperature suggested in \cite{Strasberg21} assumes that one has one has absolutely no control on systems $B$ and that the systems interacting with it will not be sensitive to the non-thermal features of $B$. Going beyond this assumption to investigate how nonequilibrium properties in $B$ can be used as resources is precisely a motivation of our approach.

 Finally, one can actually consider intermediate situations where the experimentalist (or ar system) can only access some of the nonequilibrium resources stored in the state of $B$ (e.g. it can only extract work by inducing coherent displacements of a bosonic system $B$ \cite{Maffei21}). 
In appendix D, we derive an effective second law consistent with treating only those accessible resources as work, which holds provided the initial state of $B$ does not contain any of the other (inaccessible) resources. This formulation has the operational advantage to need to track only the variation of the accessible resources, not the full state of $B$.

\section{Retrieving ideal sources of work and heat}\label{s:idealsources}

Our formalism defines work and heat exchanges between two arbitrary systems, and allows us to derive consistent expressions of the laws of thermodynamics. We finally show which limits leads to retrieve the usual cases where one of the systems is an ideal heat or work source. We first assume that system $B$
is at any time in a thermal state, namely $\rho_B(t) = w_B[\beta_B(t)]$ (with a possibly time-dependent temperature). Then, by our definitions $W_B = 0$, meaning that $B$ is a source of heat only. Note that this condition implies $\dot E_B(t) = -\beta(t)\dot S(t)$, which is consistent with a definition of ideal heat source emerging in the context of repeated interaction with units \cite{Strasberg17}. In this case, the heat can be computed from the energy variation of $B$, namely $Q_B = -\Delta E_B$ as proposed in \cite{Esposito10}. Conversely, a perfect work source is obtained when assuming that system $B$ does not become correlated with system $A$, i.e. $I_{AB}(t) = 0$ at any time $t$ or equivalently $\rho_{AB}(t) = \rho_A(t)\otimes\rho_B(t)$. It is then easy to show that system $B$ follows a unitary evolution, along which neither its entropy nor its effective temperature $\beta_B(t)$ vary, leading to $Q_B(t) = 0$ and $W_B(t) = -\Delta E_B(t)$ (see Appendix E). Moreover, system $A$ evolves under the action of effective Hamiltonian $H_A^\text{eff}(t) = H_A + \text{Tr}\{\rho_B(t)V\}$, and the work can be computed according to the broadly used formula $\dot W_B = \text{Tr}\left\{\rho_A(t)\dfrac{\partial}{\partial t}H_A^\text{eff}(t)\right\}$. The internal dynamics of $B$ is then responsible for the emergence of a time-dependent Hamiltonian for $A$, often considered to model a work source.

One motivation of our formalism is precisely to interpolate between these two extremes situations and consider deviations from ideality that naturally emerge in realistic setups.

\section{Bi-partite machines}\label{s:bipartitem}

\subsection{Carnot Bounds} 

We now turn to more specific scenarios. If we first consider that we put in contact two systems which have initially the same effective temperature, i.e. $\beta_A(0)=\beta_B(0) = \beta$. Then, by energy conservation $\Delta E_A + \Delta E_B +\Delta E_{\rm int} =0$, Eq.\eqref{Clausius} implies:
\bb
W_A +W_B-\Delta E_\text{int}(t)\leq 0,
\ee
namely, the total amount of non-thermal energy in the systems, which as seen earlier can be treated as a reserve of work, tends to decrease (up to the energy $-\Delta E_\text{int}$ which is paid when decoupling the two systems), preventing us, as expected, from building a heat engine working with a single temperature. On the other hand, if $\beta_A\neq\beta_B$, one can examine the conditions to build an autonomous, microscopic engine. Let us focus on the case of a refrigerator aiming at cooling down system $A$, i.e. $\beta_A>\beta_B$ and $Q_A\geq 0$. Eq.~\eqref{Clausius} implies:
\bb
0 \leq (\beta_A(0)-\beta_B(0))Q_A \leq \beta_B(0)\left(W_A+W_B-\Delta E_\text{int}(t)\right) .\nonumber\\\label{FridgeAB}
\ee
In the line of macroscopic thermodynamics, we see that such process is only possible if some resource is consumed: either system $A$ or $B$ provides work by reducing their energy of non-thermal nature, either energy is provided by a decrease of the coupling energy, which in turn will require an amount of work $-\Delta E_\text{int}(t) \geq 0$ to couple and then decouple the systems. 
The efficiency of conversion of those resources into cooling power $Q_A$ is upper bounded by the Carnot coefficient of performance $\eta_\text{Carnot}^\text{COP} = \beta_B(0)/(\beta_A(0)-\beta_B(0))$. The striking novelty is that the macroscopic principles ruling the performances of a refrigerator are here extended for two arbitrary quantum systems, in the absence of macroscopic thermal bath and of external work source. This is exemplified below on the case of two qubits, where it is shown that the role of the hot bath (the entropy sink) and the work source can be both played by one of the two qubits so as to cool down the other.
Conversely, one can also consider elementary engines, where work is generated within $A$ or $B$ out of a temperature gradient between the thermal backgrounds of $A$ and $B$. The efficiency of the engine is upper bounded by the Carnot efficiency $(\beta_A(0)-\beta_B(0))/\beta_A(0)\leq 1$ for $\beta_A(0)\geq \beta_B(0)$ (see plots in examples below).

\begin{figure*}
    \centering
    (a)\includegraphics[width=0.5\textwidth]{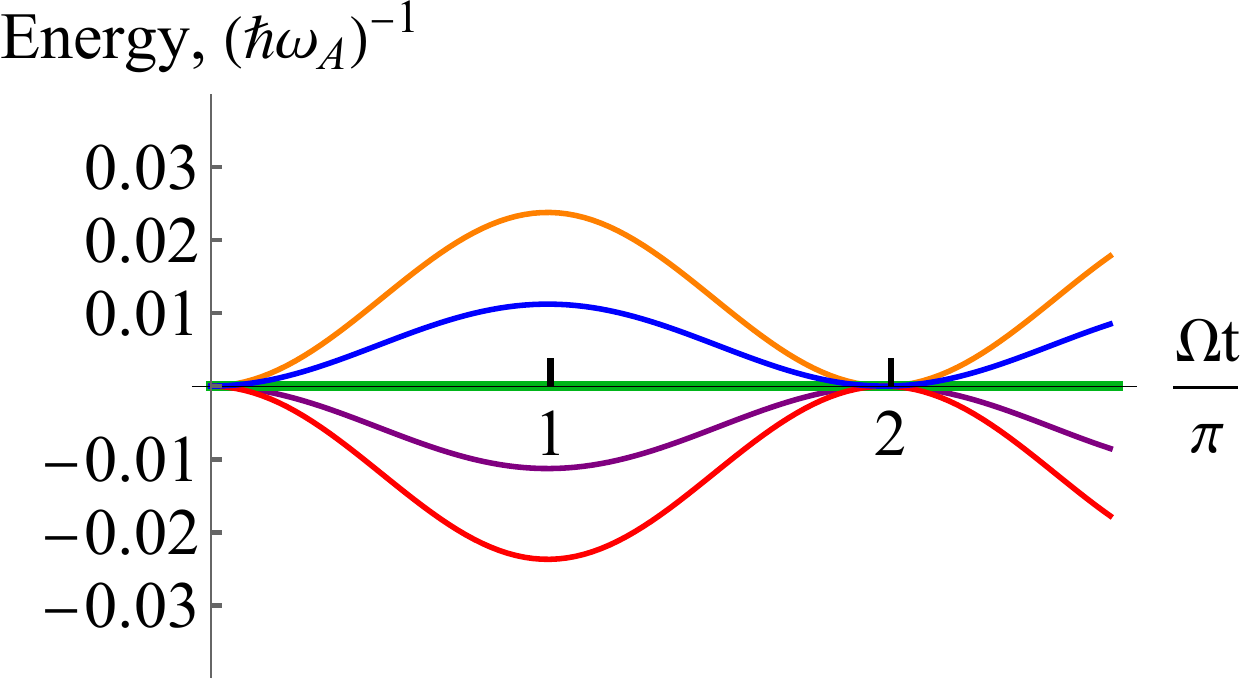} (b) \includegraphics[width=0.4\textwidth]{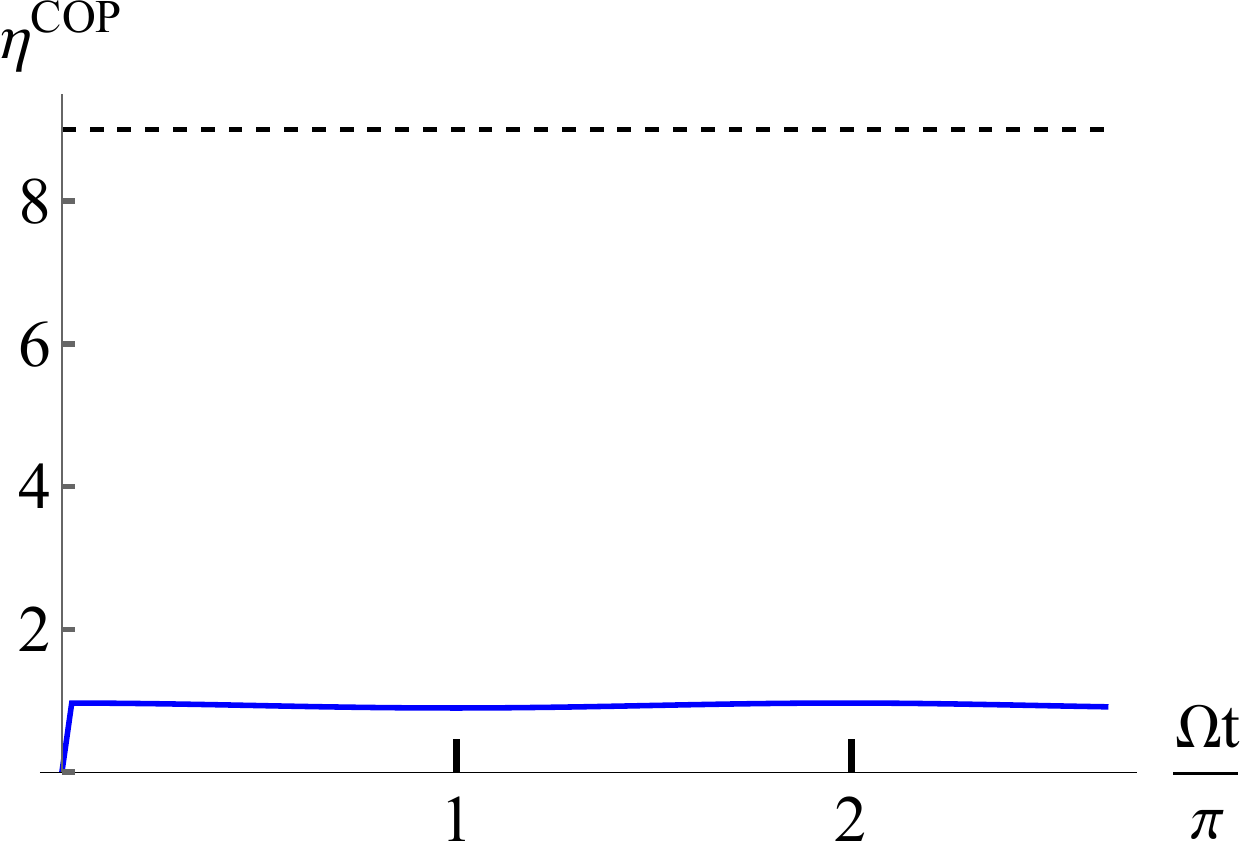}
    \caption{\textbf{Performances of the two-qubit refrigerator}. \textbf{a}: heat flow $Q_A(t)$ from qubit $A$ (blue curve), heat flow $Q_B(t)$ from qubit $B$ (red curve), work $W_A(t)$ provided by qubit $A$ (purple curve), work $W_B(t)$ provided by qubit $B$ (orange curve), and coupling energy $E_\text{int}(t)$ (green curve) in units of $\omega_A$ after the qubits interacted for a duration $t$. Time is given in unit of the frequency $\Omega = \sqrt{g^2+(\omega_B-\omega_A)^2}$ characterizing the effective strength of the coupling. \textbf{b}: coefficient of performance of the refrigerator $\eta^\text{COP}(t) = Q_A(t)/[W_A(t)+W_B(t)-\Delta E_\text{int}(t)]$ as a function of the interaction time $t$ (blue solid line). Carnot's upper bound $\eta_{\rm Carnot}^\text{COP}= \beta_B(0)/(\beta_A(0)-\beta_B(0))$ defined above is indicated by the dashed line. {\it Parameters}: $\beta_A(0)\omega_A = 2$, $\beta_B\omega_A = 1.8$, $\omega_B/\omega_A = 1.25$, $g/\omega_A =0.5$, $\phi = 0.055\pi$.}
    \label{f:2}
\end{figure*}

\subsection{Example: a refrigerator composed of two qubits}\label{s:qubitfridge}

We consider two qubits of Hamiltonians $H_j = (\omega_j/2)\sigma_z^j$, with $j=A,B$ and $\sigma^j_{x,y,z}$ are the Pauli matrices in the Hilbert space of $j$. Initially, the qubit $A$ is in a thermal state $\rho_A(0) = w_A[\beta_A(0)]$ at smaller temperature than $B$, i.e. $\beta_A(0) \geq \beta_B(0)$. The goal of the refrigerator is to extract heat from qubit $A$ and reject it into qubit $B$. While in a traditional refrigerator this would be obtained by spending work from an external drive, it can be seen from Eq.~\eqref{FridgeAB} that one can also consume non-thermal energy initially present in qubit $B$. The latter then plays the role of both the hot bath and the work source needed to build a generic heat engine. To be specific, we assume that the initial state of qubit $B$ is $\rho_B(0) = e^{-i\phi\sigma_x^B}w_B[\beta_B(0)]e^{i\phi\sigma_x^B}$, i.e., depending on the value of $\phi$, a state with coherences and/or inverted-population in the energy eigenbasis. We also consider that the two qubits are coupled via the Hamiltonian $V =  g\sigma_+^{A}\otimes\sigma_-^{B}/2 + {\rm h.c.}$, where $\sigma_{\pm} := \sigma_x \pm i \sigma_y$ are the raising and lowering qubit operators. From the solution of the Schrödinger equation for the two qubits, we computed analytically the heat flow $Q_A(t)$ provided by the qubit $A$, the work $W_B(t)$ provided by the qubit $B$, and the variation of the internal energy. Their expressions are given in Appendix F and are plotted in Fig.~\ref{f:2}\textbf{a}, showing that refrigeration indeed occurs at all times except for  $t$ equal to a multiple of  $ 2\pi/\sqrt{g^2+(\omega_B-\omega_A)^2}$. Note that for the chosen parameters, the energy of $A$ is almost constant since work is stored within $A$ simultaneously with heat being released to $B$. The refrigeration is thus characterized by a net decrease of the entropy of $A$. We also plot on panel \textbf{b} the coefficient of performance of the refrigerator $\eta(t)$ together with Carnot's bound. 
The relative low performance of the refrigerator observed in Fig.~\ref{f:2}\textbf{b} is due to the generation of correlations between $A$ and $B$ as well as the increase of the thermal distances $D(w_B[\beta_B(t)]\|w_B[\beta_B(0)])$ and $D(w_A[\beta_A(t)]\|w_A[\beta_A(0)])$, that is a consequence of the extreme smallness of the heat sources.

\subsection{Example: an engine composed of two qubits}\label{s:qubitengine}

\begin{figure*}
    \centering
    (a)\includegraphics[width=0.5\textwidth]{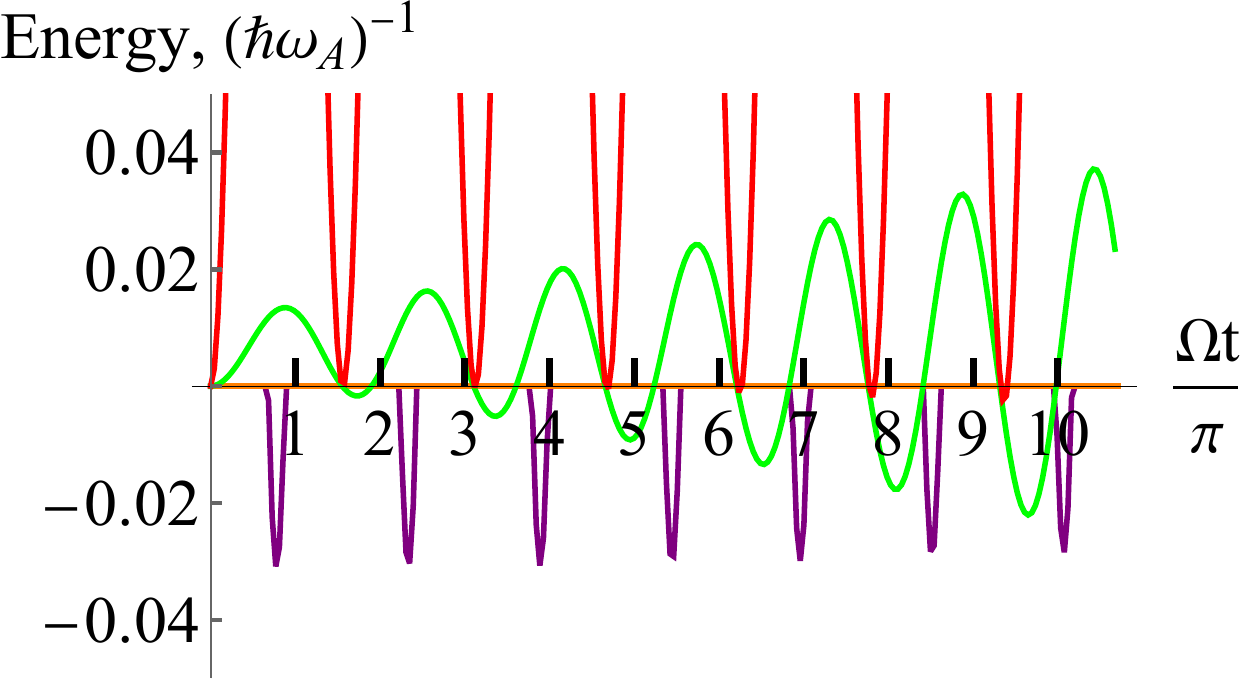}(b)\includegraphics[width=0.4\textwidth]{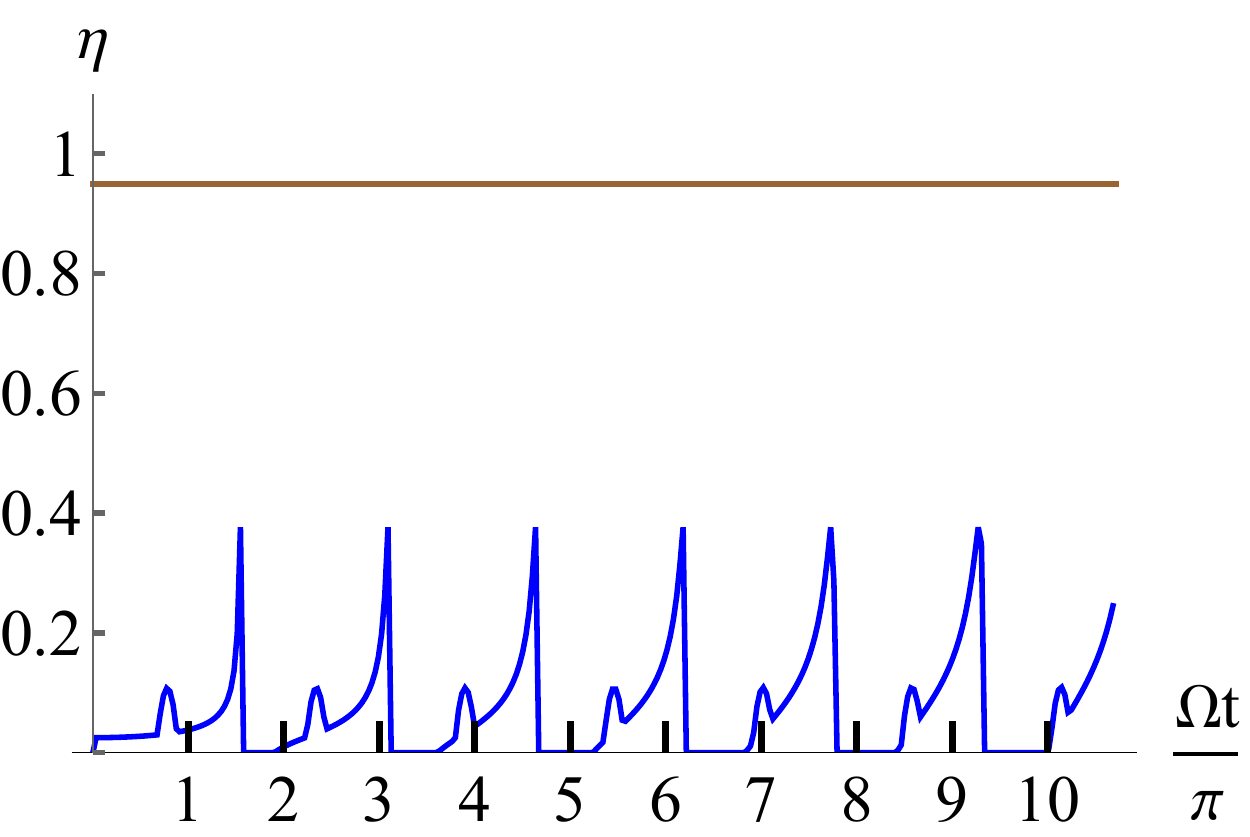}
    \caption{\textbf{Performances of the two-qubit engine}. \textbf{a}: Work $W_A(t)$ provided by $A$  (purple curve), heat $Q_B(t)$ provided by $B$ (red curve), and variation of the coupling energy $\Delta E_{\rm int}(t)$ (green curve) in units of $\omega_A$ after the qubits interacted during a time $t$. Time is given in unit of the frequency $\Omega = \sqrt{g_x^2+g_y^2+(\omega_B-\omega_A)^2}$. 
    Note that $W_B(t)$ is equal to 0 for the considered time interval, and thus is not appearing in the plot. \textbf{b}: Efficiency of the engine $\eta(t)= \frac{[-W_A -W_B + \Delta E_{\rm int}]}{Q_B} \Theta(-W_A -W_B + \Delta E_{\rm int})$, with $\Theta$ the Heaviside step function (blue curve) and Carnot's efficiency $1-\beta_B(0)/\beta_A(0)$ (brown straight line). {\it Parameters}: $\beta_A(0)\omega_A = 2$, $\beta_B(0)\omega_A = 0.1$, $\omega_B/\omega_A = 1.63$, $g_x/\omega_A =2$, $g_y/\omega_A = 0.8$.}
    \label{f:3a}
\end{figure*}

The complementary operation consists in converting heat from the hotter system into work or coupling energy. To achieve this we consider the same two qubits $A$ and $B$, but this time initialized in thermal states at temperatures $\beta_A(0)>\beta_B(0)$. The qubits are now coupled via Hamiltonian  $V =g_x \sigma_x^A\sigma_x^B/2 + g_y \sigma_y^A\sigma_y^B/2$. As shown in Fig.\ref{f:3a}, we found parameters allowing such engine to store work only in one of the two qubits (namely qubit $A$). However, part of the thermal bias is used to generate interaction energy. The efficiency of the process is plotted in Fig.\ref{f:3}(b). Note that the expected trade-off between work extraction and efficiency can also be retrieved in such machine: by changing the coupling strength, one can control the speed of the work extraction stroke (i.e. move the time at which the first peak of efficiency in Fig.\ref{f:3}(b) occurs). As numerically shown in Appendix G, a shorter stroke results in lower efficiency of conversion.

\subsection{Example: exploiting passive non-thermal states}\label{S:ExtractionPassive}

The two-qubit engine above can be extended to demonstrate the possibility to extract work from the non-thermal energy of passive states (i.e. states with zero ergotropy). To do so, we replace qubit $B$ with a $(d+2)$-level qudit chosen to have the following properties: (i) The ground and first-excited levels are separated by the frequency $\omega_B$, and are coupled to qubit $A$ with the same coupling Hamiltonian as in Section \ref{s:qubitengine} ; (ii) The other $d$-levels are assumed for simplicity to have all the same energy $\omega_2>\omega_B$ and to be uncoupled with qubit $A$; (iii) The qudit is initially in a passive state $\rho_B(0) = \sum_{k=0}^d p_k \ket{E_k}_B\bra{E_k}$, with $p_1/p_0 = e^{-\beta_B(0)\omega_B}$ for $\beta_B(0)<\beta_A(0)$, and $p_k/p_1 = e^{-\beta_2\omega_2}$, $\forall k\geq 2$, for some inverse temperature $\beta_2>0$ chosen such that the total effective initial temperature of qudit $B$ is equal to $\beta_A(0)$. We have denoted $\ket{E_k}_B$ the eigenstates of $H_B$. For the same values of $\beta_{A,B}(0)$ and $\omega_{A,B}$ as in Fig.~\eqref{f:3}, these three conditions ensure that the joint dynamics of 
the qudit $B$ and qubit $A$ will consume the only available resource, i.e. non-thermal energy in the qudit, to store work in the qubit (see Appendix H for further details).

\section{Describing fully autonomous machines}

The examples treated in Sections~\ref{s:qubitfridge} and \ref{s:qubitengine} are not fully autonomous as the coupling between the systems must be switched on or off externally to start/stop the operation of the machines. However, the present framework is well suited to analyze autonomous machines as we sketch below. 

\subsection{Steady-state engines} 

One first possible kind of autonomous machines stems from a nonequilibrium steady state between different reservoirs and work sources continuously performing a given task (e.g. conversion of heat into work) \cite{Linden2010,Levy2012,Correa2013, Gelbwaser2014,Gelbwaser2015,Mitchison2016,Mitchison2019,Latune2019}.
Treating such situations requires to consider large (or infinite) systems for $A$ and $B$. In this spirit, it has already been suggested that infinite baths in squeezed and/or displaced states behaves has source of work \cite{Niedenzu2016,Manzano2016,Manzano2018, Rodrigues2019}. When combined with (exact or approximate) methods to express the systems' dynamics, our results provide a unified framework to analyze such engines and investigate resource conversion mechanisms \cite{Bernhardt22}.\\

\subsection{Autonomous piston} \label{s:autop}

Another class of autonomous machines involve additional degrees of freedom whose dynamics will impose the targeted engine schedule (which can be referred to as a clock or a piston) \cite{Elouard15,Monsel18,Strasberg21c}. In principle, any non-autonomous setup involving a time-dependent Hamiltonian can be mimicked this way \cite{Brandao13,Alhambra16}. We provide in Appendix I an example of ideal clock model, based on the motion of massive particle, able to induce autonomous switching of the interaction between two systems, and e.g. make the machines of section \ref{s:bipartitem} fully autonomous. The ideal clock turns out to be an ideal work source (as defined in Section \ref{s:idealsources}) and the work it provides matches the energy variation associated with switching on and off the interaction between systems $A$ and $B$. Such quantum clock is closely related to ideal work storage models used in resource-theoretic formulations of quantum thermodynamics \cite{Brandao13,Alhambra16}. More realistic autonomous thermodynamic transformations of finite duration can also be investigated from quantum scattering models \cite{Jacob21,Jacob22}.

\subsection{Second law for $N$ systems}

It is likely that any useful autonomous machine will be made of more than two interacting systems. This is the case for instance when the driving protocol ruling the interaction between two sources is rules by the dynamics of a clock as proposed in the previous section~\ref{s:autop}, or when a working medium is coupled to multiple baths. We prove in Appendix J that our main results Eq.~\eqref{2ndLaw_AB} and \eqref{Clausius} can be extended to a set of more than two systems initially uncorrelated and put in contact at a time $t=0$, leading to:
\bb
\Delta S_j -\sum_{i\neq j}\beta_j(0)Q_j(t) \geq 0\label{2ndLawN}
\ee
and
\bb
-\sum_j  \beta_j(0)Q_j(t) \geq 0,\label{ClausiusN}
\ee
where $Q_i(t) = -\Delta E_i^\text{th}(t)$ and $\beta_i(t)$ are defined analogously as before for each system $i$. Eqs.~\eqref{2ndLawN}-\eqref{ClausiusN} can be used to investigate realistic setups of quantum heat engines, where some systems play the role of non-ideal hybrid work and heat sources, while others play the role of working media.

\section{Generalizations}\label{s:Generalizations}

\subsection{Initial correlations}

The violation of Eqs.~\eqref{2ndLawN}-\eqref{ClausiusN} means (i) that the systems $S_i$ are not evolving unitarily, which can be useful for instance to detect the presence of noise in the setup \cite{Uzdin18}; or (ii) the systems are initially correlated. Building up on this last possibility, it is a known fact that the consumption of initial correlations is a resource to extract work or invert the sign of natural heat flows \cite{Sagawa12,Micadei19} (e.g. it captures the case of Maxwell demons). We can therefore refine the inequalities to include such situations, obtaining
\bb
-\sum_j  \beta_j(0)Q_j(t) \geq \Delta I_\text{tot}(t),\label{ClausiusNCorr}
\ee
where $ I_\text{tot}(t) = D(\rho_\text{tot}(t)\| \bigotimes_{j=1}^N \rho_j(t))$ quantifies correlations between the $N$ systems at time t, and $\Delta I_\text{tot}(t)$ is the {\it variation} of correlations, not necessarily positive. Another approach to take into account these initial correlations was proposed in \cite{Uzdin18,Uzdin19}.

\subsection{Bounds involving the time-dependent temperatures}

Finally, it can be noticed that as the effective temperatures $\beta_i(t)$ of the systems generally depend on time, their thermal part can be seen as finite-size heat baths. In particular, from Eqs.~\eqref{2ndLawN}-\eqref{ClausiusN} one can derive tighter inequalities using the instantaneous temperatures to quantify the entropy flows:
\bb\label{tighterEP}
\Delta S_j - \sum_{i\neq j}\int_0^tdt'\beta_i(t') \dot Q_i(t')\geq 0,\nonumber\\
{\rm and} ~~~ -\int_0^tdt' \sum_j \beta_j(t') \dot Q_j(t') \geq 0,
\ee
where $\dot Q_j(t) = -\dot E_j^\text{th}(t)$ is the heat current associated to $Q_j(t)$.

Note that inequalities of the same form have been derived in \cite{Strasberg21,Strasberg21b}, but from a different approach treating asymetrically small systems and bath. In particular, different notions of nonequilibrium bath temperature (computed from the bath's energy) and of heat (identified with the whole change of energy of systems treated as baths) are used (see also discussion in Section \ref{s:accessefft}); this leading to expressions of entropy production which are positive only for some class of initial bath states. In contrast, we here provide a symmetric description, quantifying all nonequilibrium resources in the systems, therefore leading to a positive entropy production whatever the initial states of the system.
A subsequent generalization of the Carnot bound is given in Appendix K.

\section{Conclusion and outlook} 
We have demonstrated a version of the second law of thermodynamics valid for any set of quantum systems, initially in arbitrary uncorrelated states, and evolving autonomously under their joint unitary evolution. 

The initial entropy of each system leads to the emergence of an effective temperature and an amount of thermal energy, the variation of which plays the role of heat. The energy beyond the thermal energy can be instead used as a resource to decrease entropy, and its variation is therefore assimilated to work. Moreover, this quantity is reminiscent of the notion of ergotropy, already considered as a promising candidate to extend the notion of work to the quantum domain. We also showed that these notions of heat and work become equivalent to earlier well-accepted expressions in the case of systems behaving as ideal work and heat sources. We illustrated the reach of our results by demonstrating the possibility to design refrigerators and heat engines based on microscopic quantum systems each able to combine simultaneously the role of work and heat sources. With our universally valid notions of work and heat, these microscopic engines then follow laws similar to macroscopic thermodynamics, with efficiencies complying with Carnot's bounds.

The consequences of our results are multifaceted and yet to be explored. On one hand, they ground recent attempts to refine the splitting between work and heat in a fully quantum description. On the other hand, they allow one to understand the performances of realistic quantum devices, in which no such things as perfect heat sources or work sources exist, and determine which microscopic properties lead to devices which are close to, or conversely deviate from, these ideal models. 
While our results hold in principle at any scale, a practical limitation comes from the challenge to quantify and access all the nonequilibrium resources (the generalized ergotropy) stored in a large quantum system. This difficulty can be addressed within our formalism by defining easily accessible resources, and treating the others as part of the heat, which leads to an effectively coarse-grained description, still finer than the paradigm of pure heat and work sources (see Section \ref{s:accessefft} and Appendix D). This modular methodology can directly be combined with the theoretical frameworks \cite{Esposito09,Hartmann17} and experiments \cite{Eichler11,Wenniger22}  which provide promising partial access to quantum baths dynamics.
When instead focusing on machines made of elementary systems, our formalism opens an avenue towards the design of extremely compact quantum engines, whose performances constitutes a largely uncharted territory.

\subsection*{Acknowledgements} 
This work was supported by funding from French National Research Agency (ANR) under grant ANR-20-ERC9-0010 (project QSTEAM). The authors thank Chris Jarzynski, Benjamin Huard, Patrice Camati, Ariane Soret, Reouven Asouly and Philipp Strasberg for helpful discussions and comments.

\section*{Appendix A: Interpretation of ${\cal E}^\infty$ as the preparation work cost}

\setcounter{equation}{0}
\makeatletter
\renewcommand{\theequation}{A\arabic{equation}}
\makeatother

We are interested in this section in the minimum amount of work ones needs to prepare an arbitrary non-equilibrium state $\rho_A^0$ of a system $A$. For reference, the same conclusion as the one derived below was obtained in \cite{Bera19}.

We assume that the system $A$ of Hamiltonian $H_A$ is initially in an equilibrium state $w_A[\beta, H_A]:= \frac{e^{-\beta H_A}}{\text{Tr}\{e^{-\beta H_A}\}}$ with a bath at inverse temperature $\beta$. Assuming we can perform arbitrary driving and quenches, one reversible protocol to reach $\rho_A^0$ is as follows.

\begin{itemize}
 \item Perform an isothermal reversible quasi-static driving $H_A\rightarrow \tilde H_A$ such that the final state is our target state, $ w_A[\beta,\tilde H_A] = \rho_A^0$.  This can always be done, at least in theory, for arbitrary initial state $\rho_A^0$. It can be shown as follows. One expresses $\rho_A^0$ in its diagonal form,
\bb
\rho_A^0 := \sum_{l=1}^L r_l |r_l\>\<r_l|,
\ee
and
\bb
H_A = \sum_{k=1}^K e_k |e_k\>\<e_k|,
\ee 
with $1\leq L \leq K \leq +\infty$. For $l \in [1;L]$, we introduce pseudo-energies $E_l$ as 
\bb
E_l := - \frac{1}{\beta}(\ln r_l + c),
\ee
 where $c$ is a free constant one can use to choose the energy origin. Then, if $\rho_A^0$ is full rank, meaning if $L=K$, we define $\tilde H_A$ as 
 \bb
 H_A := \sum_{l=1}^L E_L |r_l\>\<r_l|.
 \ee
However, if $\rho_A^0$ is not full rank $L<K$, we define $\tilde H_A$ as
 \bb
 H_A := \sum_{l=1}^L E_L |r_l\>\<r_l| + \sum_{l=L+1}^K E_l |e_l\>\<e_l|,
 \ee
 where $E_l = E_{l'} \gg 1/\beta$ for all $l, l' \geq L+1$. With this choice, one can verify that 
 \bb
 w_A[\beta,\tilde H_A] := e^{-\beta \tilde H_A} /{\rm Tr} [e^{-\beta \tilde H_A}] = \rho_A^0,
 \ee
 for the full-rank situation. This identity becomes only approximate in the non-full-rank situation, but the approximation is exponentially good for large $E_l$, $l>L$. The work involved in this reversible quasi-static driving is given by the variation of equilibrium free energy $F[w[\beta, H]] := {\rm Tr} [\rho H] - \frac{1}{\beta}S[\rho]$,
 \bb
 W_{\rm quasi-static} =  F[w_A[\beta, \tilde H_A]]  - F[w_A[\beta, H_A]] .
 \ee 
 
 \item Switch off the bath interaction.
 \item Perform a quench $\tilde H_A \rightarrow H_A$ to come back to the initial Hamiltonian. The work involved in the quench is 
  \bb
 W_{\rm quench} = {\rm Tr}[\rho_A^0( H_A -\tilde H_A)],
 \ee
 since the quench does not change the state of $A$ for occurring on a timescale much smaller than the system evolution time.
 \end{itemize}  
 Then, the total work invested to prepare $\rho_A^0$, is 
 \bb
 W_\beta &=& W_{\rm quasi-static} + W_{\rm quench}\nn\\
 &=& {\cal F}_{\beta,H_A}[\rho_A^0]-  F[w_A[\beta, H_A]]  
 \ee
 where ${\cal F}_{\beta,H_A}[\rho_A^0] = {\rm Tr}[\rho_A^0 H_A] - \frac{1}{\beta}S[\rho_A]$ denotes the non-equilibrium free-energy. Interestingly, one can show that the overall work $W$ is related to the relative entropy between $\rho_A^0$ and $w_A[\beta,H_A]$,
 \bb
 W_\beta = \frac{1}{\beta} D(\rho_A^0\|w_A[\beta,H_A]),
 \ee
 which is always positive, as expected: one always has to spend work to prepare a non-equilibrium state. Since the above protocol is reversible, it guarantees to be the one with the smallest amount of work to invest, for fixed $\beta$. Then, what is the value of $\beta$ which minimizes $W_\beta$?
 The answer is given by taking the derivative of $W_\beta$ with respect to $\beta$. We obtain
 \bb
 \frac{1}{\partial \beta} W_\beta &=& \frac{1}{\partial \beta} F[w_A[\beta, \tilde H_A]]-\frac{1}{\partial \beta} F[w_A[\beta, H_A]]   \nn\\
 &=& \frac{1}{\beta^2}\left\{S[\rho_A^0] - S[w_A[\beta, H_A]] \right\}.
 \ee
Since $S[w_A[\beta, H_A]] $ is a monotonic decreasing function of $\beta$, we deduce that $W_\beta$ is a monotonic decreasing function of $\beta$ on $[0;\beta^0]$ and a monotonic increasing function of $[\beta^0;+\infty[$, where $\beta^0$ denotes the inverse temperature such that $S[w_A[\beta^0,H_A]] = S[\rho_A^0]$. Consequently, the minimum amount of work to prepare $\rho_A^0$ is $W_{\beta^0}$. By noticing that $W_{\beta^0} = {\rm Tr}[\rho_A^0H_A] - {\rm Tr}[w_A[\beta^0,H_A]H_A] = {\cal E}^\infty_A$, the ``generalized" ergotropy \cite{Alicki13} introduced in the main text, one concludes that $ {\cal E}^\infty_A$ represents the minimal amount of work needed to prepare the state $\rho_A^0$ out of a thermal bath.

\section*{Appendix B: Negativity of the entropy production defined from the splitting ergotropy - passive energy}

\setcounter{equation}{0}
\makeatletter
\renewcommand{\theequation}{B\arabic{equation}}
\makeatother

One natural choice widely used for autonomous systems is to identify the variation of ergotropy of a system as work. If one assumes that the variation of ergotropy corresponds to the whole exchange of work, the second law of thermodynamics should take the form,
\bb\label{2ndlawerg}
\sigma_{A,\text{erg}}(t) := \Delta S_A + \beta_B(0)\left(\Delta E_B - \Delta {\cal E}_B\right) \geq 0.
\ee
From Eq.~\eqref{2ndLaw_1}, one finds that:
\bb
\sigma_{A,\text{erg}}(t) &=& I_{AB}(t)+  D[\pi_B(t)|w_B[\beta_B(0)]]\nonumber\\
&&- D[\pi_B(0)|w_B[\beta_B(0)]].
\ee
We now show that the quantity on the right-hand side can become negative, or equivalently that the free energy of system $A$ can be increased by more than the ergotropy consumed. We first note that in the case where system $B$ is a qubit, $E_B -{\cal W}_B =E_B^\text{th}$, such that $\sigma_{A,\text{erg}} = \sigma_A \geq 0$. Our counterexample must therefore involve a system of dimension at least $3$. We consider the case where systems $A$ and $B$ are two identical qutrits of Hamiltonians
\bb
H_j =  \sum_{k=0}^2 \omega_k \ket{k}_j\bra{k}_j,\quad j=A,B 
\ee
initialized in state $\rho_{AB}(0) = \rho_A(0)\otimes\rho_B(0)$, with $\rho_A(0) = w_A[\beta_B(0)]$ a thermal state at temperature $\beta(0)$ and $\rho_B(0) =  \sum_{k=0}^2 p_{B,k} \ket{k}_B\bra{k}_B$. We further assume that $p_{B,2}\leq p_{B,1}\leq p_{B,0}$ (that is $\rho_B(0)$ is a passive state and ${\cal W}_B(0) = 0$), but \emph{non-thermal}, that is, there is no positive real number $\beta$ such that $p_{B,i}/p_{B,j}= e^{-\beta(\omega_i-\omega_j)}$ for all couples $(i,j)\in \llbracket 0,2\rrbracket^2$. Such a non-thermal passive state can be simply built by choosing $p_{B,1}\propto e^{-\beta_1 \omega_1}p_{B,0}$ and $p_{B,2} \propto e^{-\beta_2 \omega_2}p_{B,0}$ with two different positive numbers $\beta_1\neq\beta_2$. These conditions imply that $E_B(0) - E_B^\text{th}(0)\neq 0$. As before, we denote $\beta_B(0)$ the inverse temperature of the thermal state which has the same entropy as $\rho_B(0)$. Finally, we choose a coupling Hamiltonian implementing a swap of the two qutrit states, namely:
\bb
V = g \sum_{k\neq l}\ket{kl}_{AB}\bra{lk}. 
\ee
It is straightforward to show that for $t=t_\text{SWAP}$ verifying $gt_\text{SWAP} = \pi/2$, we have $\rho_{AB}(t_\text{SWAP}) = \rho_B(0)\otimes\rho_A(0)$, that is the states of qutrits $A$ and $B$ are swapped. This means that the state of qutrit $B$ is replaced with a thermal state which has the same entropy, and therefore still no ergotropy ${\cal W}_B(0)$, however its energy has decreased by $E_B(t_\text{SWAP})-E_B(0) = E_B^\text{th}(0)-E_B(0)  <0$. Moreover, the entropy of qubit $A$ did not change:
\bb
S_A(t) &=& S[\rho_A(t_\text{SWAP})]
= S[\rho_B(0)]\nonumber\\ &=& S[w_B[\beta_B(0)]] = S_A(0).
\ee
Finally, we have $\sigma_A < 0$. The qutrit swap allows for a spontaneous diminution of the passive amount of energy $E_B - {\cal W}_B$ in qubit $B$.
Similarly, the variation of the free energy of system $A$ is equal to its variation of internal energy as $\Delta S_A = 0$, that is $\Delta E_A(t_\text{SWAP}) = -\Delta E_B(t_\text{SWAP}) \geq 0$: the free energy of qubit $A$ was increased without consuming any ergotropy. 

Consequently, identifying work with the variation of ergotropy may lead to underestimation of entropy production. More generally, as the present example allows for a spontaneous decrease of the passive amount of energy $E_B - {\cal W}_B$ of $B$ without any entropy variation of the entropy of $A$, it also demonstrates that there is no other choice of effective temperature that one could inject in Eq.~\eqref{2ndlawerg} to ensure its positivity: this demonstrates that the notion of ergotropy is insufficient to capture all the energy that can be assimilated as work. One can note however that under special assumptions, e.g. restricting the possible initial state of system $B$ to unitary-transformed thermal states, i.e. $\rho_B(0) = U_0 w_B[\beta_B(0)]U_0^\dagger$, one obtains 
\bb  
\Delta S_A(t) + \beta_B(0)\Delta {\cal E}_A(t) \geq 0,
\ee
in agreement with the results of \cite{Uzdin18,Niedenzu18}. Note that is inequality is less tight than Eq.~\eqref{2ndLaw_AB} in general as system $B$ can still end up in a state that contains more non-thermal energy than ergotropy.

\section*{Appendix C: Symmetric formulation of the 2nd law}

\setcounter{equation}{0}
\makeatletter
\renewcommand{\theequation}{C\arabic{equation}}
\makeatother

To go from Eq.~\eqref{2ndLaw_AB} to inequality \eqref{Clausius} of main text, we use the definitions of $w_A[\beta_A(t)]$:
\bb
\Delta S_A &=&  S[w_A[\beta_A(t)]]-S[w_A[\beta_A(0)]]\nonumber\\
&=& S[w_A[\beta_A(t)]]+\text{Tr}\{w_A[\beta_A(t)]\log w_A[\beta_A(0)]\}\nonumber\\
&&-\text{Tr}\{w_A[\beta_A(t)]\log w_A[\beta_j(0)]\}-S[w_A[\beta_A(0)]]\nonumber\\
&&- \text{Tr}\{w_A[\beta_A(0)]\log w_j[\beta_A(0)]\}\nonumber\\
&&+ \text{Tr}\{w_A[\beta_A(0)]\log w_j[\beta_A(0)]\}\nonumber\\
&=& -D(w_A[\beta_A(t)\| w_A[\beta_A(0)])\nonumber\\
&&-\text{Tr}\{w_A[\beta_A(t)]\log w_A[\beta_A(0)]\}\nonumber\\
&&+ \text{Tr}\{w_A[\beta_A(0)]\log w_A[\beta_A(0)]\}.
\ee
This finally leads to the identity:
\bb
\Delta S_A  &=& -D(w_A[\beta_A(t)\| w_A[\beta_A(0)])-\beta_A(0)Q_A(t),\quad\;\;\;\label{DSQ}\ee
valid for any quantum system.\\

\section*{Appendix D: Partial control and extractable work}\label{app:pctrl}
\setcounter{equation}{0}
\makeatletter
\renewcommand{\theequation}{D\arabic{equation}}
\makeatother

In this section, we focus for simplicity on the case of two systems $A$ and $B$, while the results below can straightforwardly be extended for $N$ systems. 
To formalize the idea of partial access to the degree of freedom of $B$ introduced in the main text \ref{s:accessefft}, we assume that we can act on $B$ only via a family of unitaries $U[\alpha]$ which consume one type of resources that are considered accessible. For instance, for a  bosonic bath, $U[\alpha]$ could be displacement operators, as energy stored in a coherent displacement can be efficiently re-used as work by driving a system quasi-resonantly \cite{Monsel20}. We then assume that the initial state of $B$ has the form $\rho_B(0) = U^\dagger[\alpha_0] w_B[\beta(0)] U[\alpha_0]$ to show that
\begin{align}\label{eq:2ndLawacc}
    \sigma_\alpha &:= \Delta S_A -\beta(0){\cal Q}_B^{\alpha}\nonumber\\
    &= I_{AB}(t) + D(U[\alpha_t]\rho_B(t) U^\dagger[\alpha_t]\| w_B[\beta(0)])\geq 0.
\end{align}
Above, 
\begin{align}
    {\cal Q}_B^{\alpha} &= -\text{Tr}\{H_B\left(U[\alpha_t]\rho_B(t) U^\dagger[\alpha_t]-U[\alpha_0]\rho_B(0) U^\dagger[\alpha_0]\right)\}\nonumber\\
    &=-\text{Tr}\{H_B\left(U[\alpha_t]\rho_B(t) U^\dagger[\alpha_t]-w_B[\beta(0)]\right)\}
\end{align}
gathers the heat $Q_B$ provided by system $B$ (as defined in main text) and all the non-thermal energy that cannot be extracted by one of the unitaries $U[\alpha]$. $U[\alpha_t]$ is the unitary of the family that allows to extract work from the final state of $B$. Conversely,\\
\begin{align}
    {\cal W}_B^\text{acc} &= \text{Tr}\{H_B\left(\rho_B(0)-U[\alpha_0]\rho_B(0) U^\dagger[\alpha_0]\right)\}\nonumber\\
  &-\text{Tr}\{H_B\left(\rho_B(t)-U[\alpha_t]\rho_B(t) U^\dagger[\alpha_t]\right)\},
\end{align}
is the work provided by $B$ via a variation of the accessible resources only. 

\noindent\emph{Proof}.-- Defining $\tilde\rho_B(t) = U[\alpha_t]\rho_B(t) U^\dagger[\alpha_t]$, we write 
\begin{eqnarray}
 D(\tilde\rho_f(t)\|w_f[\beta(0)]) &=& \text{Tr}\{\tilde\rho_f(t)\log \tilde\rho_f(t)\} \nonumber\\&&-\text{Tr}\{\tilde\rho_f(t)\log w_f[\beta(0)]\}\nonumber\\
 &=& \text{Tr}\{\rho_f(t)\log \rho_f(t)\} \nonumber\\&&-\text{Tr}\{\tilde\rho_f(t)\log w_f[\beta(0)]\}\nonumber\\
 &=& \text{Tr}\{\rho_f(t)\log \rho_f(t)\}\nonumber\\&& - \text{Tr}\{\rho_f(0)\log \rho_f(0)\}\nonumber\\
 &&+ \text{Tr}\{\rho_f(0)\log \rho_f(0)\} \nonumber\\&&-\text{Tr}\{\tilde\rho_f(t)\log w_f[\beta(0)]\}\nonumber\\
 &=& -\Delta S_f + \text{Tr}\{w_f[\beta(0)]\log w_f[\beta(0)]\} \nonumber\\&&-\text{Tr}\{\tilde\rho_f(t)\log w_f[\beta(0)]\}\nonumber\\
  &=& -\Delta S_f\nonumber\\&& -\beta(0) \text{Tr}\{H_B\left(w_f[\beta(0)]-\tilde\rho_f(t)\right)\}.\nonumber\\
 \end{eqnarray}
 We then inject this equation for $\Delta S_B$ into the identity $\Delta S_A + \Delta S_B = I_{AB}$ to derive $\sigma_\alpha$. \\
 
 Note that Eq.~\eqref{eq:2ndLawacc} is less tight than Eq.~\eqref{2ndLaw_AB}. However, $\sigma_\alpha$ can become negative if the initial state of $B$ contains resources which are not accessible via $U(\alpha)$.
  Nevertheless, the coarse-grained description associated with $\sigma_\alpha$ does not requires access to the entropy of $B$, while still constituting a finer description of the energy exchanged than Eq.~\eqref{2ndLaw_1}.

\section*{Appendix E: Ideal work source}
\setcounter{equation}{0}
\makeatletter
\renewcommand{\theequation}{E\arabic{equation}}
\makeatother

We assume that the unitary evolution of $A$ and $B$ is such that at any time, $I_{AB}(t) = 0$ and $\rho_{AB}(t) = \rho_A(t)\otimes\rho_B(t)$. The Liouville-Von-Neumann evolution equation reads:
\bb  
\dot\rho_{AB}(t) = -i[H_A+V+H_B,\rho_A(t)\otimes\rho_B(t)].
\ee
Taking partial trace over one or the other system, we obtain:
\bb  
\dot\rho_{A}(t) &=& -i[H_A^\text{eff}(t),\rho_A(t)]\nonumber\\
\dot\rho_{B}(t) &=& -i[H_B^\text{eff}(t),\rho_B(t)]\label{eq:evolB},
\ee
where $H_A^\text{eff}(t) = H_A + \text{Tr}_B\{V\rho_{B}(t)\}$ (respectively $H_B^\text{eff}(t) = H_B + \text{Tr}_A\{V\rho_{A}(t)\}$) is an effective Hamiltonian acting on system $A$ (resp. $B$), where $\text{Tr}_i$ denotes the partial trace over the operator space of system $i$. We can see that both systems undergo a unitary evolution due to their interaction. As a consequence, the entropy of $B$, and therefore its effective inverse temperatures $\beta_B(t)$ are conserved. This leads to $Q_B(t) = 0$ and $W_B(t) = -\Delta E_B(t)$. We now use Eq.~\eqref{eq:evolB} to express the work provided by system $B$. We have
\bb
W_B(t) &=& -\int_0^t dt' \text{Tr}\{H_B\left(-i[H_B^\text{eff}(t),\rho_B(t)]\right)\}\nonumber\\
&=& i\int_0^t dt' \text{Tr}\{[H_B,H_B+\text{Tr}_A\{V\rho_A(t)\}]\rho_B(t)\}\nonumber\\
&=& i\int_0^t dt' \text{Tr}\{[H_B,\text{Tr}_A\{V\rho_A(t)\}]\rho_B(t)\}\nonumber\\
&=& i\int_0^t dt' \text{Tr}\{[H_B^\text{eff}(t),\text{Tr}_A\{V\rho_A(t)\}]\rho_B(t)\}\nonumber\\
\nonumber\\
&=&i\int_0^t dt'\text{Tr}\{[ H_B^\text{eff}(t),V]\rho_A(t)\otimes\rho_B(t)\}\nonumber\\
&=& -i\int_0^t dt'\text{Tr}\Big\{ \text{Tr}_B\{V[H_B^\text{eff}(t),\rho_B(t)]\}\rho_A(t)\Big\}\nonumber\\
&=& \int_0^t dt'\text{Tr}\Big\{  \frac{\partial H_A^\text{eff}(t)}{\partial t}\rho_A(t)\Big\}.
\ee
We finally note that in the case of only two subsystems, the condition that $B$ is an ideal work source for $A$ implies automatically that $A$ is also an ideal work source for $B$, such that the entropy of both system is constant. As soon as three or more systems are coupled however, the entropy of system $A$ may vary due to interactions with other energy sources which are not ideal work sources, even though the entropy of $B$ remain constant. \\

\section*{Appendix F: Autonomous two-qubit  refrigerator}

\setcounter{equation}{0}
\makeatletter
\renewcommand{\theequation}{F\arabic{equation}}
\makeatother

For a qubit system characterized by state $\rho_j(t)$, the Von Neumann entropy depends only of the parameter $r_j(t) = \sqrt{\text{Tr}\{\sigma^j_x\rho_i(t)\}^2+\text{Tr}\{\sigma^j_y\rho_i(t)\}^2+\text{Tr}\{\sigma_z^j\rho_i(t)\}^2}$, namely:
\bb
S[\rho_j(t)] &=& -\frac{1+r_j(t)}{2}\log\left(\frac{1+r_j(t)}{2}\right)\nonumber\\
&&-\frac{1-r_j(t)}{2}\log\left(\frac{1-r_j(t)}{2}\right).
\ee
Comparison with the entropy of a qubit thermal state $w_j[\beta_j] = e^{-\beta_j\omega_j\sigma_z^j/2}/Z_i$ allows one to identify the effective temperature $\beta_j(t) = \log((1+r_j(t))/(1-r_j(t))/\omega_j$. Consequently, we can compute the thermal energy $E_j^\text{th}(t) = - r_j(t)\omega_j/2$, and therefore the work and heat provided by qubit $i$ between $t=0$ and $t$:
\bb
Q_j &=& \omega_i \Delta r_j(t)/2\\
W_j &=& -\omega_i (\Delta r_j(t)+\Delta z_j(t))/2,
\ee
where $z_j(t)= \text{Tr}\{\sigma_z^j\rho_j(t)\}$. Assuming the initial states $\rho_A(0) = w_A[\beta_A(0)]$ and $\rho_B = e^{-i\phi\sigma_x^B}w_B[\beta_B(0)]e^{i\phi\sigma_x^B}$, we can propagate the two-qubit state according to $\rho_{AB}(t) = e^{-iH_{AB}t}\rho_A(0)\otimes\rho_B(0)U^{iH_{AB}t}$ and compute:
\begin{widetext}
\bb
x_A(t) &=& 2\frac{g}{\Omega}\cos\phi\sin\phi\tanh\frac{\omega_A\beta_A(0)}{2}\tanh\frac{\omega_B\beta_B(0)}{2}\cos\frac{(\omega_A+\omega_B)t}{2}\sin \frac{\Omega t}{2} \\
y_A(t) &=& - 2\frac{g}{\Omega}\cos\phi\sin\phi\tanh\frac{\omega_A\beta_A(0)}{2}\tanh\frac{\omega_B\beta_B(0)}{2}\sin\frac{(\omega_A+\omega_B)t}{2}\sin \frac{\Omega t}{2} \\
z_A(t) &=& \frac{1}{4\Omega^2}\left\{-[2g^2(1+\cos \Omega t) + 4(\omega_A-\omega_B)^2]\tanh\frac{\omega_A\beta_A(0)}{2}-2g^2(1-\cos\Omega t)\cos 2\phi\tanh\frac{\omega_B\beta_B(0)}{2}\right\}\\
x_B(t) &=& \frac{2}{\Omega}\cos\phi\sin\phi \tanh\frac{\omega_B\beta_B(0)}{2}\left[-(\omega_A-\omega_B)\cos\frac{(\omega_A+\omega_B)t}{2}\sin\frac{\Omega t}{2} +\Omega \sin\frac{(\omega_A+\omega_B)t}{2}\cos\frac{\Omega t}{2} \right]\\
y_B(t) &=& \frac{2}{\Omega}\cos\phi\sin\phi \tanh\frac{\omega_B\beta_B(0)}{2}\left[(\omega_A-\omega_B)\sin\frac{(\omega_A+\omega_B)t}{2}\sin\frac{\Omega t}{2} +\Omega \cos\frac{(\omega_A+\omega_B)t}{2}\cos\frac{\Omega t}{2} \right]\\
z_B(t) &=&  \frac{1}{4\Omega^2}\left\{-2g^2(1-\cos \Omega t)\tanh\frac{\omega_A\beta_A(0)}{2} -[2g^2(1+\cos\Omega t)+4(\omega_A-\omega_B)^2]\cos 2\phi\tanh\frac{\omega_B\beta_B(0)}{2}\right\}.\\
\ee
 Moreover:
\bb
\Delta E_\text{int}(t) = -\frac{g^2(\omega_A-\omega_B)(1-\cos\Omega t)}{2\Omega^2}\frac{(e^{\omega_A\beta_A(0)}-e^{\omega_B\beta_B(0)})\cos^2\phi +(e^{\omega_A\beta_A(0)+\omega_B\beta_B(0)}-1)\sin^2\phi}{(e^{\omega_A\beta_A(0)}+1)(e^{\omega_B\beta_B(0)}+1)}
\ee
\end{widetext}
We have introduced the parameter:
\bb
 \Omega = \sqrt{g^2+(\omega_A-\omega_B)^2}.
\ee\\

\section*{Appendix G: Trade-off velocity versus efficiency}
\setcounter{equation}{0}
\makeatletter
\renewcommand{\theequation}{G\arabic{equation}}
\makeatother

\begin{figure*}
    \centering
    (a)\includegraphics[width=0.45\textwidth]{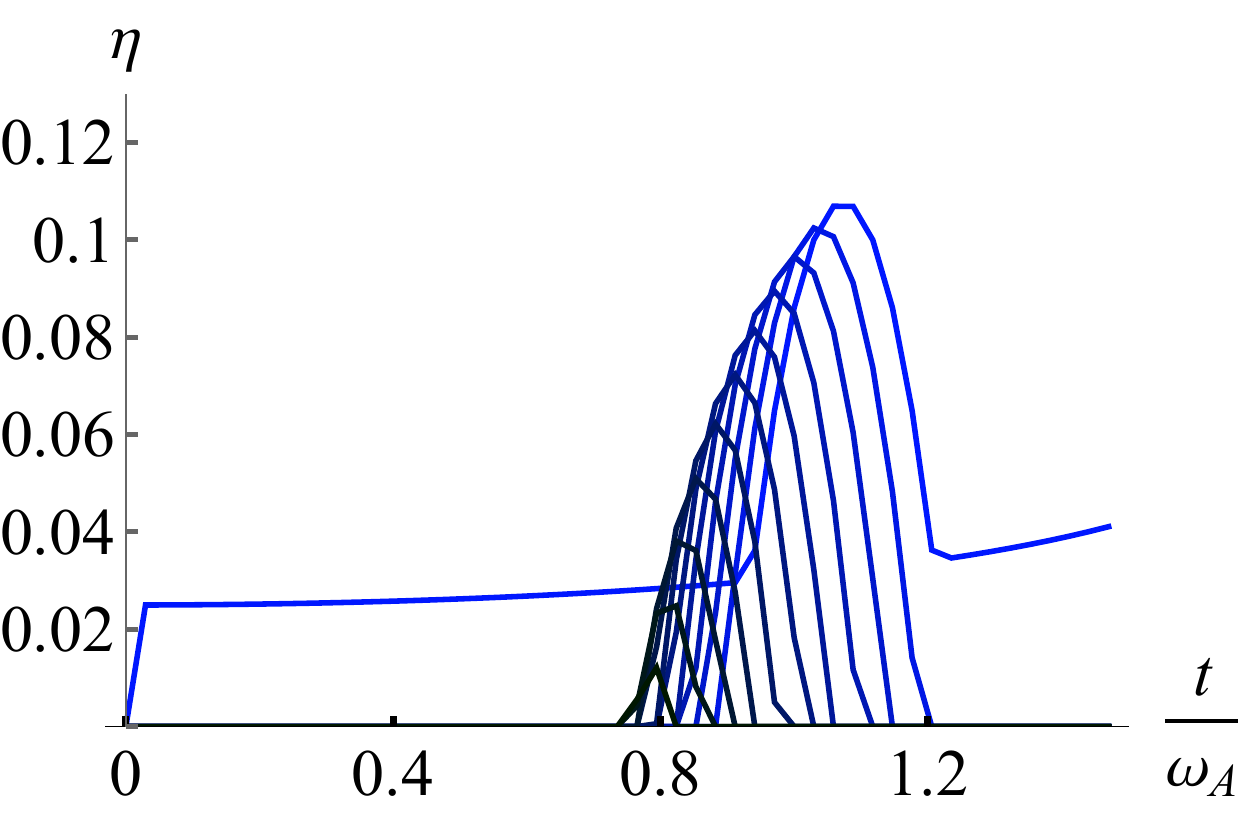}
    \caption{\textbf{Trade-off velocity of the work production versus efficiency}. Plots of the efficiency of the autonomous two-qubit engine against the time (in unit of $\omega_A$), for coupling strength varying from $g_x/\omega_A=2$ (lighter curve) to $g_x/\omega_A=3$ (darker curve). The other parameters are the same as in the previous figure, namely $\beta_A(0)\omega_A = 2$, $\beta_Br\omega_A = 0.1$, $\omega_B/\omega_A = 1.63$, and $g_y/\omega_A = 0.8$.}
    \label{f:3}
\end{figure*}

In this section we aim to briefly illustrate that the heat (and potentially the work) provided by the system $B$ does depend on the ``trajectory" of $A$. One general consequence is that faster processes are usually more dissipative/irreversible and less efficient. 
In Fig. \eqref{f:3}, we plot the efficiency of the autonomous two-qubit engine considered in section \ref{s:qubitengine} against the time evolution (in unit of $\omega_A$) for increasing coupling strength. The lighter curve corresponds to $g_x/\omega_A = 2$, while the darker curve corresponds to $g_x/\omega_A = 3$. Each curve in-between corresponds to an increment of $\delta g_x/\omega_A = 0.1$ with respect to its lighter neighbor. We observe the expected behavior: faster processes are less efficient. This illustrates, in particular, that the heat provided by $B$ does depend on the trajectory of $A$, as announced. 

\section*{Appendix H: Extracting work from a non-thermal passive state}

\setcounter{equation}{0}
\makeatletter
\renewcommand{\theequation}{H\arabic{equation}}
\makeatother

\subsection*{Conditions to reproduce the two-qubit work extraction}

We consider a bi-partite machine made up of a qubit $A$ and a $(d+2)$-level qudit $B$. $A$ is initially in a thermal state $w_A[\beta_A(0)]$, while $B$ is initially in a non-thermal passive state $\rho_B(0)=\sum_k p_k \ket{E_k}_B\bra{E_k}$, where $\{\ket{E_k}_B\}$ is the eigenbasis of $H_B$, with an effective temperature matching that of qubit $A$, i.e. $\beta_A(0)$. As qudit $B$ is not initially in a thermal state, the matching temperatures \emph{do not} imply equilibrium and the non-thermal energy of qudit $B$ can be consumed to generate heat and work flows between the systems. We wish to choose qudit $B$ such that its two lower levels behave as the second qubit of the two-qubit engine presented in the main text. We therefore choose an energy splitting $\omega_B$ between these two levels, and a population ratio $p_1/p_0 = e^{-\beta_B(0)\omega_B}$ with $\beta_B(0)<\beta_A(0)$. Moreover, we couple these two levels of the qudit to qubit $A$ with the same coupling as for the two-qubit engine. We further assume that the other states of qudit $B$ are not coupled to qubit $A$.

When this conditions are fulfilled, the joint dynamics of the qubit and of the two first levels of the qudit exactly match that of the two-qubit engine presented in the main text. To prove this statement, we write the qudit Hamiltonian as a block matrix:
\bb 
H_B =  \left(\begin{array}{cc}
    \omega_2 \idop_2\otimes\idop_{d} & 0 \\
     0 & \omega_B \idop_2\otimes(1+\sigma_z)/2
\end{array}\right)
\ee
Here $\idop_{n}$ denotes the identity in dimension $n$ and tensor products are always denoted with the space of qubit $A$ on the left. We assume that only qubit $B'$ is coupled to qubit $A$:
\bb 
V' =  \left(\begin{array}{cc}
    0 & 0 \\
     0 & V\end{array}\right),
\ee
with $V=\frac{g_x}{2}\sigma_x\otimes\sigma_x+\frac{g_y}{2}\sigma_y\otimes\sigma_y$.
Denoting $H_A= (\omega_A/2)\sigma_z\otimes \idop_{d+2}$ and $H =H_A+H_B+V'$, the total unitary evolution is:
\bb 
U_{AB} = \left(\begin{array}{cc}
    e^{-i\omega_2t}\idop_2\otimes\idop_{d}  & 0 \\
     0 & U_\text{2x2}
\end{array}\right),
\ee
with $U_\text{2x2}= e^{-iH_\text{2x2}t}$ the two-qubit reduced evolution ruled by $H_\text{2x2}= (\omega_A/2)\sigma_z\otimes\idop_2 + (\omega_B/2)\idop_2\otimes\sigma_z + V$, which is exactly the dynamics considered in Section \eqref{s:qubitengine}.

\subsection*{Initial state of qudit $B$}

We now examine the condition of existence of an admissible initial passive state of $B$ whose effective temperature matches that of qubit $A$ while fulfilling $p_1/p_0 = e^{\beta_B(0)\omega_B}$ with $\beta_B(0)<\beta_A(0)$. The desired initial state of  $B$ is:
\bb
\rho_{B}(0) = \frac{1}{\cal N}\left(\begin{array}{cc}
    e^{-\beta_2\omega_2} \idop_d & 0 \\
     0 & e^{-\beta_B\omega_B(1+\sigma_z)/2}
\end{array}\right).
\ee

We first note that $S_B(0)$ is a decreasing function of $\beta_2$ and reaches its minimum value $\underset{\beta_2\to\infty}{\text{Lim}}S_B(0) = S_{B'}(0)$, i.e. the initial entropy of the qubit $B'$ encoded in the two lower level of $B$. Therefore a necessary condition for the existence of an inverse temperature $\beta_2$ such that the effective temperature of $B$ is $\beta_A(0)$ is that $S_{B'}(0)= S[w_{B'}[\beta_B(0)]]\leq S[w_B[\beta_A(0)]]$.

To have an inverse effective temperature $\beta_A(0)$, the qudit initial entropy must be equal to
\begin{widetext}
\bb
S[w_B[\beta_A(0)]]
&=& -\text{Tr}\left\{\frac{1}{{\cal N}_3}\left(\begin{array}{cc}
    e^{-\beta_A(0)\omega_2} \idop_d \log(e^{-\beta_A(0)\omega_2}/{\cal N}_3) & 0 \\
     0 & \begin{array}{cc}  e^{-\beta_A(0)\omega_B}\log(e^{-\beta_A(0)\omega_B}/{\cal N}_3) & 0\\ 0& \log(1/{\cal N}_3)  \end{array}
\end{array}\right)\right\}\nonumber\\
&=& \frac{{\cal N}_2}{{\cal N}_3}S[w_{B'}[\beta_A(0)]]+\frac{{\cal N}_3-{\cal N}_2}{{\cal N}_3}\log(d) +S_\text{Sh}\left(\frac{{\cal N}_2}{{\cal N}_3}\right),
\ee
\end{widetext}
with ${\cal N}_3 = 1+d e^{-\beta_A(0)\omega_2}+e^{-\beta_A(0)\omega_B}$, ${\cal N}_2 = 1+e^{-\beta_A(0)\omega_B}$ and $S_\text{Sh}(x) = -x\log x -(1-x)log(1-x)$ the Shannon entropy. We see that $S[w_B[\beta_A(0)]]$ is constituted of the average of $S[w_{B'}[\beta_A(0)]< S[w_{B'}[\beta_B]]$ and $\log(d)$, plus a positive term given by a Shannon entropy associated with the binary distribution $\left\{\frac{N_2}{N_3}, 1-\frac{N_2}{N_3}\right\}$. When $de^{-\beta_2\omega_2}\ll 1$ (small values of $d$ or of $\beta_A\omega_2$), then $S[w_B[\beta_A(0)]]$ is close to $S[w_{B'}[\beta_A(0)]$ which is smaller than $S[w_{B'}[\beta_B(0)]$  and therefore the equality cannot be satisfied. This result let us foresee the existence of constraints beyond Eq.~\eqref{2ndLaw_AB} on the practical ability to consume passive non-thermal energy as a resource.
On the other hand, for sufficiently large value of $d$, $S[w_{B}[\beta_A(0)]>S[\omega_{B'}[\beta_B(0)]]$. This is for instance clearly true when $d\to\infty$ such that $S[w_{B}[\beta_A(0)] \to \log d$ while $S_{B'}(0)\leq \log 2$. Thus, by increasing enough the dimension of the qudit, one can always find a passive non-thermal state reproducing the two-qubit engine.\\

To reproduce the engine of Section \eqref{s:qubitengine}, $d=5$, $\omega_2 = 1.7\omega_A$ and $\beta_2 = 75.97$ are sufficient.

\section*{Appendix I: Clock model for autonomous driving}
\setcounter{equation}{0}
\makeatletter
\renewcommand{\theequation}{I\arabic{equation}}
\makeatother

In this section, we present one idealized model showing the possibility to generate any effective time-dependent Hamiltonian from the interaction with an auxiliary system initialized out of equilibrium.

Specifically, we consider that the interaction Hamiltonian between systems $A$ and $B$ is controlled by another degree of freedom, denoted C, and hereafter called ``clock''. 
For instance, A and B could be the internal degrees of freedom of two particles, interacting via a distance-dependent coupling, while C is the motional degree of one of the particles. Then, one can prepare C in a state localized around a large distance to the other particle (ensuring vanishing initial interaction between A and B), but with an average initial velocity ensuring that the subsequent dynamics will bring the two particle close enough to reach significative interaction for some duration $\tau$.

In an ideal setup \cite{Alhambra16,Gisin18}, the clock is described by a Hamiltonian $H_C = v\hat p_C$ (analogous to a free photon), with $\hat p_C$ the momentum operator of C, and initialized in a eigenstate of the position $\hat q_C \ket{q_0}_C = q_0 \ket{q_0}_C$. The Hamiltonian of the total autonomous system read:
\bb 
H &=& H_A+H_B+H_C +V_{ABC}\\
V_{ABC} &=& G(\hat q_C)V_{AB},
\ee
where $V_{AB}$ is a Hamiltonian coupling systems A and B only, and $G(q)$ is a continuous function with finite support (typically a gate function). The coupling between systems A and B is switched on when the clock's spaial wavepacket has an overlap with the region verifying $G(q)\neq 0$, that we assume to be centered around $q=0$ without loss of generality. We assume that $q_0 <0$ lies outside this region. One can understand the autonomous switching of the coupling in the interaction picture with respect to the free clock's Hamiltonian, where dynamics is governed by Hamiltonian:

\bb  
H^I(t) = H_A +H_B + G(\hat q_C + vt)V_{AB}\label{d:HIABC}
\ee

Finally, the reduced dynamics of system A and B is effectively ruled by an effective Hamiltonian with time-dependant coupling autonomously switching on and off provided the wavepacket of the clock remain much localized in position space with respect to the typical scale of variation $G(q)$. This typically coincides with the scattering limit ensured by a large initial mean energy and energy uncertainty of the clock \cite{Jacob21,Jacob22,Rogers22}

Taking into account the initial state of the clock, the full system dynamics obeys
\begin{widetext}
\bb  
\rho_{ABC}(t) &=& e^{-iHt}\left(\rho_{AB}(0)\otimes\ket{q_0}\bra{q_0}\right)e^{iHt}\\
&=&  e^{-iH_C t} {\cal T}e^{-i\int_0^t dt'H^I(t')}\left(\rho_{AB}(0)\otimes\ket{q_0}\bra{q_0} \right) \left({\cal T}e^{-i\int_0^t dt'H^I(t')}\right)^\dagger e^{iH_Ct}\\
&=&  {\cal T}e^{-i\int_0^t dt'H^\text{eff}_{q_0}(t')}\rho_{AB}(0) \left({\cal T}e^{-i\int_0^t dt'H^\text{eff}_{q_0}(t')}\right)^\dagger \otimes e^{-iH_C t}\ket{q_0}\bra{q_0} e^{iH_Ct}\\
&=&  {\cal T}e^{-i\int_0^t dt'H^\text{eff}_{q_0}(t')}\rho_{AB}(0) \left({\cal T}e^{-i\int_0^t dt'H^\text{eff}_{q_0}(t')}\right)^\dagger \otimes \ket{q_0+vt}\bra{q_0+vt} 
\ee
\end{widetext}
where 
\bb
H^\text{eff}_{q_0}(t) = H_A +H_B + G(q_0 + vt)V_{AB}
\ee
is the effective time-dependent Hamiltonian controlling the reduced dynamics of system A and B. We note that system C is always in a factorized state with system A and B. To go from the second to the third line, we have used that the interaction picture Hamiltonian Eq.\eqref{d:HIABC} preserves the clock's state $\ket{q_0}\bra{q_0}$.

From Section \ref{s:idealsources} of main text and Appendix G, we see that the clock behave as an ideal work source and the work it provides verifies:
\bb
W_C(t) &=& \int_0^t dt'\text{Tr}\left\{\frac{\partial H^\text{eff}_{q_0}(t')}{\partial t'}  \rho_{AB}(t')\right\}\nonumber\\
&=& \int_0^t dt'\frac{\partial}{\partial t'} \text{Tr}\left\{ H^\text{eff}_{q_0}(t')\rho_{AB}(t')\right\}\\
&=& G(q_0 + vt)\text{Tr}\{V_{AB}\rho_{AB}(t)\}\nonumber\\
&&- G(q_0)\text{Tr}\{V_{AB}\rho_{AB}(0)\}\nonumber\\
&&  \Delta E_A + \Delta E_B.
\ee
That is, any difference between the total energy of system $A+B$ between initial and final time, including the interaction energy, corresponds to work performed by the clock. Additionally, the entropy of the clock remains constant and it does not provide any heat. 

The case of autonomous switching of the coupling corresponds to $G(q_0)=G(q_0+vt)=0$ and $G(q)\simeq 1$ approximately constant over the interval $(q_1,q_2)$ where it is non-zero. The rise of $G(q)$ from $0$ to $1$ is assumed to be fast with respect to the dynamics of $A$ and $B$. Moreover, we assume that the energies of  $A$ and $B$ are each conserved when the two system don't interact, i.e. outside the time interval $t'\in (t_1,t_2)$, with $t_1=\frac{q_1-q_0}{v}$ and $t_2=\frac{q_2-q_0}{v}$.
From these assumptions, we obtain:
\bb
W_C(t) &=&  \int_{t_1^-}^{t_1^+} dt'\text{Tr}\left\{\frac{\partial H^\text{eff}_{q_0}(t')}{\partial t'}  \rho_{AB}(t')\right\}\nonumber\\
&+& \int_{t_2^-}^{t_2^+} dt'\text{Tr}\left\{\frac{\partial H^\text{eff}_{q_0}(t')}{\partial t'}  \rho_{AB}(t')\right\}\nonumber\\
&=& \moy{V_{AB}(t_1^+)} - \moy{V_{AB}(t_2^-)}.
\ee
Thus, the energy required to switch on and off the coupling corresponds to the work provided by the clock.\\

While preparing a particle in an infinite energy state such as a position eigenstate is obviously unrealistic, the situation above is well approximated by considering a finite wavepacket of small position (large momentum) variance. The Hamiltonian $H_C = v\hat p_C$ can be approximated by a free massive particle of large mean initial impulsion $p_0$ such that any variation of its kinetic energy caused by the interaction can be linearized, i.e.:
\bb 
\frac{\hat p_C^2}{2m} \sim \frac{p_0^2}{2m} + \frac{\hat p_C - p_0}{m} + O(\hat p_C - p_0)^2,
\ee 
where $m$ is the particle's mass. We finally note that such an ideal clock model is similar to the work storage model used in Refs.~\cite{Alhambra16}.
\\

\section*{Appendix J: Autonomous second law for $N$ systems}
\setcounter{equation}{0}
\makeatletter
\renewcommand{\theequation}{J\arabic{equation}}
\makeatother

We here extend the formulation of the second law to the case of $N$ quantum systems initially in a factorized state $\rho_\text{tot}(0) = \bigotimes_{i=1}^N \rho_i(0)$. The total Hamiltonian reads $H_\text{tot} = \sum_i H_i + V$, where $V$ contains all the coupling terms.
We introduce:
\bb
I_\text{tot}(t) = D(\rho_\text{tot}(t)\| \bigotimes_{i=1}^N \rho_i(t)) \geq 0,
\ee
 which verifies $I_\text{tot}(0)=0$.
We can then express the variation of the Von Neumann entropy of system $j\in \llbracket 1,N\rrbracket$ between time $t=0$ and 
$t$:
\bb
\Delta S_j = I_\text{tot}(t) - \sum_{i\neq j}\Delta S_i.\label{DSj}
\ee
Analogously to the case of two systems, we define for each system $i$:
\bb
w_i[\beta_i(t)] &=& \frac{e^{-\beta_i(t)H_i}}{Z_i(t)},\\
Z_i(t) &=& \text{Tr}\{e^{-\beta_i(t)H_i}\},\\
E_i^\text{th}(t) &=& \text{Tr}\{H_i w_i[\beta_i(t)]\},\nonumber\\
Q_i(t) &=& -E_i^\text{th}(t)+E_i^\text{th}(0),
\ee
where $\beta_i(0)$ is the solution of
\bb
S[w_i[\beta_i(t)]] = S[\rho_i(0)].
\ee
We then add on both side of Eq.~\eqref{DSj} the quantity $ -\sum_{i\neq j} \beta_i(0)Q_i(t)$ and use that

\bb
&&-\beta_i(0)Q_i(t) - ( S[\rho_i(t)]-S[\rho_i(0)])\nonumber\\
&=& -\text{Tr}\{(w_i[\beta_i(t)]-w_i[\beta_i(0)])\log w_i[\beta_i(0)]\}\nonumber\\
&& - S[w_i[\beta_i(t)]] + S[w_i[\beta_i(0)]]\nonumber\\
&=& D(w_i[\beta_i(t)]\|w_i[\beta_i(0)]),
\ee
to finally get 
\bb 
&&\Delta S_j -\sum_{i\neq j}\beta_i(0)Q_i(t) =  \nonumber\\
&& I_\text{tot}(t)+ \sum_{i\neq j} D(w_i[\beta_i(t)]\|w_i[\beta_i(0)]),\label{A:2ndLawN}
\ee
which implies inequality \eqref{2ndLawN} of main text. To demonstrate  inequality \eqref{ClausiusN} of main text, we rewrite the variation the entropy of system $j$ using the identity Eq.~\eqref{DSQ} and inject it into Eq.~\eqref{A:2ndLawN} to get:
\bb
-\sum_i \beta_i(0) Q_i(t) = I_\text{tot}+\sum_i D(w_i[\beta_i(t)]\|w_i[\beta_i(0)]),\;\;\;\;\;
\ee
which in turn implies inequality \eqref{ClausiusN} of main text.

\section*{Appendix K: Inequalities involving the time-dependent temperature}\label{app:K}
\setcounter{equation}{0}
\makeatletter
\renewcommand{\theequation}{K\arabic{equation}}
\makeatother

\begin{figure}
    \centering
    (a)\includegraphics[width=0.4\textwidth]{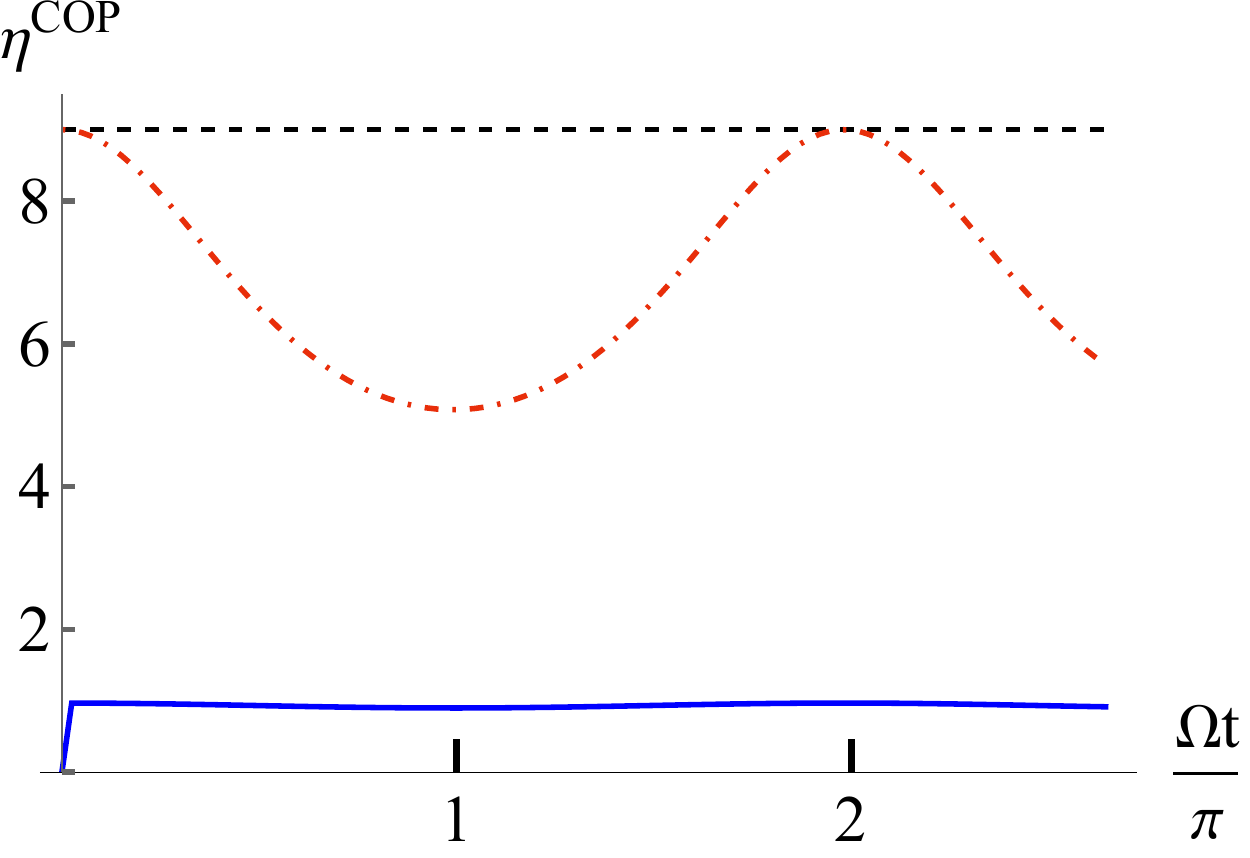}\\(b)\includegraphics[width=0.4\textwidth]{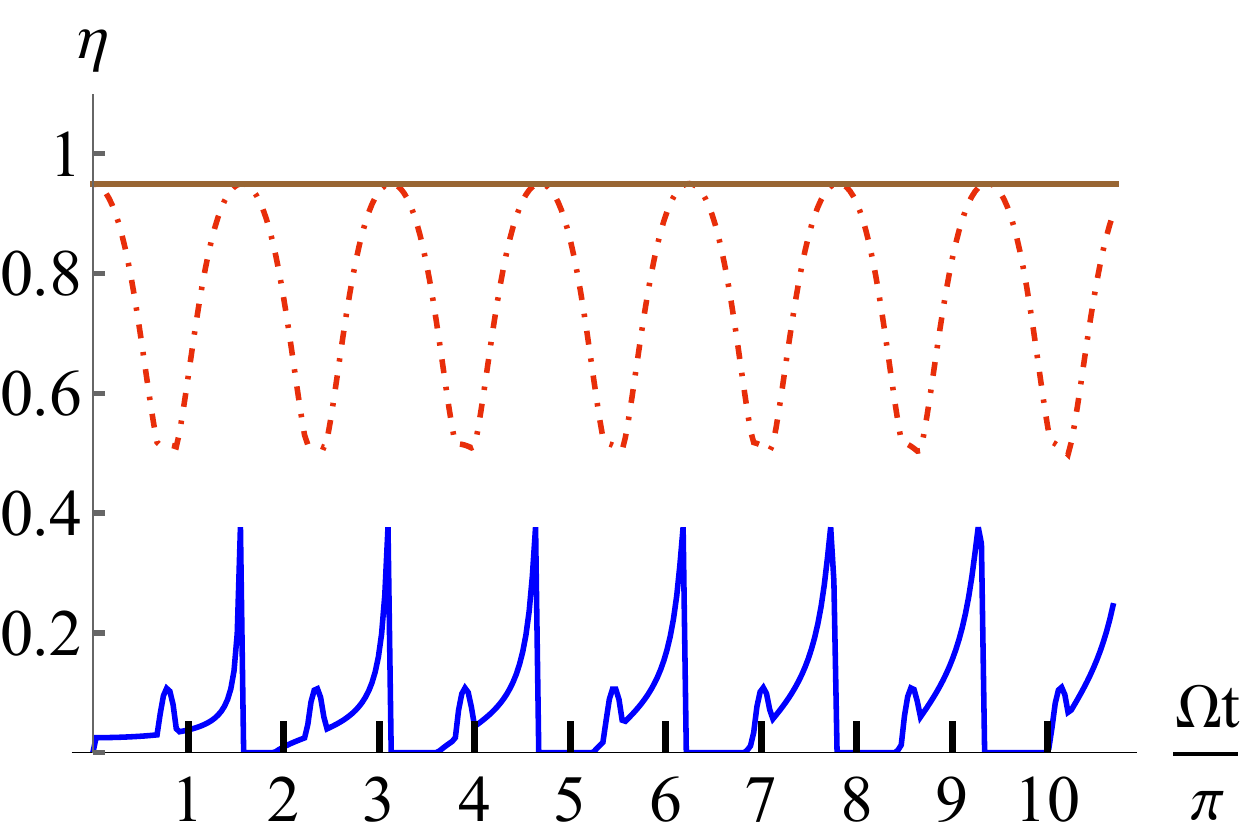}
    \caption{(a)\textbf{Performances of the two-qubit refrigerator}. Plot of the efficiency of the refrigerator presented in Fig. \ref{f:2} (b) of the main text, with the additional refined upper bound $\overline{ \eta_\text{Carnot}^\text{COP}}$ introduced in \eqref{eq:K7} in dotted-dashed line. (b)\textbf{Performances of the two-qubit engine}. Plot of the efficiency of the refrigerator presented in Fig. \ref{f:3a} (b) of the main text, with the additional refined upper bound $\overline{ \eta_\text{Carnot}}$ introduced in \eqref{eq:K9} in dotted-dashed line.}
    \label{f:5}
\end{figure}

The first line of Eq.~\eqref{tighterEP} in the main text can be seen as a tighter entropy production from which one can derive tighter upper bounds for the efficiency of engines or refrigerators. An intuitive way to see that is by introducing ``average inverse temperatures" for each system $j$ as
\bb\label{eq:K1}
\bar\beta_j \Delta E_j^{\rm th} = \int_0^t du \beta_j(u)\dot E_j^{\rm th}(u),
\ee
which verifies:
\bb \label{eq:boundbj}
\text{min}[\beta_j(0),\beta_j(t)] \leq \bar \beta_j \leq \text{max}[\beta_j(0),\beta_j(t)].
\ee
The above property can be shown as follows. 
Firstly,  let's assume, for now only, that $E_j^{\rm th}(u)$ is monotonic from $0$ to $t$. Then, one can make a change of variable  $u \rightarrow E_j^{\rm th}$ in the integral of Eq. \eqref{eq:K1}. 
This leads to 
\bb
\int_0^t du \beta_j(u)\dot E_j^{\rm th}(u) 
&=& \int_{E_j^{\rm th}(0)}^{E_j^{\rm th}(t)} dE_j^{\rm th}\beta_j[E_j^{\rm th}].
\ee
By continuity of the function $\beta_j(E_j^{\rm th})$, and since $\beta_j$ is a monotonic function of $E_j^\text{th}$, there exists $\bar \beta_j \in \big[\beta_j[E_j^{\rm th}(0)];\beta_j[E_j^{\rm th}(t)]\big]$, if $E_j^\text{th}(u)$ is increasing on  $[0;t]$ (or $\bar \beta_j \in \big[\beta_j[E_j^{\rm th}(t)];\beta_j[E_j^{\rm th}(0)]\big]$ if $E_j^{\rm th}(u)$ is decreasing on $[0;t]$), such that 
\bb
\int_{E_j^{\rm th}(0)}^{E_j^{\rm th}(t)} dE_j^{\rm th}\beta_j[E_j^{\rm th}] = \bar\beta_j [E_j^{\rm th}(t) - E_j^{\rm th}(0)].
\ee
If now $E_j^{\rm th}(u)$ is not monotonic on $[0;t]$, one can decompose the interval in smaller intervals $[t_n;t_{n+1}]$ on which $E_j^{\rm th}(u)$ is monotonic. Then, applying the above transformation of each interval $[t_n;t_{n+1}]$ we have,
\bb
\int_0^t du \beta_j(u)\dot E_j^{\rm th}(u) &=& \sum_n\int_{t_n}^{t_{n+1}} du \beta_j[E_j^{\rm th}(u)]\dot E_j^{\rm th}(u)\nn\\
&=&\sum_n \int_{E_j^{\rm th}(t_n)}^{E_j^{\rm th}(t_{n+1})} dE_j^{\rm th}\beta_j[E_j^{\rm th}]\nn\\
&=& \int_{E_j^{\rm th}(0)}^{E_j^{\rm th}(t)} dE_j^{\rm th}\beta_j[E_j^{\rm th}].
\ee
Then, repeating the same argument of continuity and monoticity of $\beta_j[E_j^\text{th}]$, we obtain,
\bb
\int_0^t du \beta_j(u)\dot E_j^{\rm th}(u) &=& \bar\beta_j [E_j^{\rm th}(t) - E_j^{\rm th}(0)],
\ee
with $\bar\beta_j$ in $[E_j^{\rm th}(0);E_j^{\rm th}(t)]$ or $[E_j^{\rm th}(t);E_j^{\rm th}(0)]$ depending whether $E_j^{\rm th}(t)\leq E_j^{\rm th}(0)$ or $E_j^{\rm th}(t) \geq E_j^{\rm th}(0)$. This corresponds to the statement \eqref{eq:boundbj}.\\

Furthermore, coming back to the two-body refrigerator introduced in section \ref{s:qubitfridge} of the main text, the use of the first line of Eq.~\eqref{tighterEP} implies
\bb
\eta^\text{COP}(t) &=& \frac{Q_A(t)}{W_A(t)+W_B(t)-\Delta E_\text{int}(t)} \nn\\
&\leq& \overline{ \eta_\text{Carnot}^\text{COP}} := \frac{\bar\beta_B}{\bar\beta_A -\bar\beta_B}.\label{eq:K7}
\ee
Additionally, one can show 
\bb
\overline{ \eta_\text{Carnot}^\text{COP}} \leq  \eta_\text{Carnot}^\text{COP} = \frac{\beta_B(0)}{\beta_A(0) -\beta_B(0)}.
\ee
In Fig.~\ref{f:5} (a) we plot the efficiency of the refrigerator as in Fig.\ref{f:2} of the main text, but adding the refined upper bound $\overline{\eta_\text{Carnot}^\text{COP}}$. The remaining distance between the actual efficiency $\eta^\text{COP}$ and $\overline{\eta_\text{Carnot}^\text{COP}}$ is solely due to the generation of correlation $I_{AB}(t)$ between $A$ and $B$. 

Similarly, for the two-body engine introduced in section \ref{s:qubitengine}, we have
\bb
\eta(t) &=& \frac{-W_A(t)-W_B(t)+\Delta E_\text{int}(t)}{Q_B(t)} \nn\\
&\leq& \overline{ \eta_\text{Carnot}} :=1 -  \frac{\bar\beta_B}{\bar\beta_A} \leq  \eta_\text{Carnot} =1 -  \frac{\beta_B(0)}{\beta_A(0)}.\nn\\\label{eq:K9}
\ee
The corresponding plot is presented in Fig.~\ref{f:5} (b).

\end{document}